\documentclass[twocolumn]{aastex61}

\usepackage{amsmath}
\usepackage{tablefootnote}
\usepackage{natbib}
\shorttitle{The SFR-M$_*$ Correlation Extends to Low Mass at High Redshift}
\shortauthors{Iyer et al.}

\begin{document}

\title{The SFR-M$_*$ Correlation Extends to Low Mass at High Redshift}

\correspondingauthor{Kartheik Iyer}
\email{kgi1@physics.rutgers.edu}

\author[0000-0001-9298-3523]{Kartheik Iyer}
\affiliation{Department of Physics and Astronomy, Rutgers, The State University of New Jersey, 136 Frelinghuysen Road, Piscataway, NJ 08854-8019 USA}

\author[0000-0003-1530-8713]{Eric Gawiser}
\affiliation{Department of Physics and Astronomy, Rutgers, The State University of New Jersey, 136 Frelinghuysen Road, Piscataway, NJ 08854-8019 USA}
\affiliation{Center for Computational Astrophysics, Flatiron Institute, 162 5th Ave, New York, NY 10010, USA}

\author{Romeel \replaced{Dave}{Dav\'{e}}}
\affiliation{Institute for Astronomy, Royal Observatory, Edinburgh EH9 3HJ, UK}
\affiliation{Department of Physics and Astronomy, University of the Western Cape, Bellville, Cape Town 7535, South Africa}

\author{Philip Davis}
\affiliation{Rutgers Discovery Informatics Institute, Rutgers, The State University of New Jersey, 96 Frelinghuysen Road, Piscataway, NJ 08854-8019 USA}

\author{Steven L. Finkelstein}
\affiliation{Department of Astronomy, The University of Texas at Austin, Austin, TX 78712, USA}

\author{Dritan Kodra}
\affiliation{Department of Physics and Astronomy and PITT PACC, University of Pittsburgh, Pittsburgh, PA 15260, USA}

\author[0000-0002-6610-2048]{Anton \added{M. }Koekemoer}
\affiliation{Space Telescope Science Institute, 3700 San Martin Drive, Baltimore, MD 21218, USA}

\author{Peter Kurczynski}
\affiliation{Department of Physics and Astronomy, Rutgers, The State University of New Jersey, 136 Frelinghuysen Road, Piscataway, NJ 08854-8019 USA}

\author{Jeffery A. Newman}
\affiliation{Department of Physics and Astronomy and PITT PACC, University of Pittsburgh, Pittsburgh, PA 15260, USA}

\author{Camilla Pacifici}
\affiliation{Space Telescope Science Institute, 3700 San Martin Drive, Baltimore, MD 21218, USA}

\author{Rachel \added{S.} Somerville}
\affiliation{Department of Physics and Astronomy, Rutgers, The State University of New Jersey, 136 Frelinghuysen Road, Piscataway, NJ 08854-8019 USA}
\affiliation{Center for Computational Astrophysics, Flatiron Institute, 162 5th Ave, New York, NY 10010, USA}
\nocollaboration










\begin{abstract}

To achieve a fuller understanding of galaxy evolution, SED fitting can be used to recover quantities beyond stellar masses (M$_*$) and star formation rates (SFRs). We use Star Formation Histories (SFHs) reconstructed via the Dense Basis method of Iyer \& Gawiser (2017) for a sample of $17,873$ galaxies at $0.5<z<6$ in the CANDELS GOODS-S field to study the nature and evolution of the SFR-M$_*$ correlation. The reconstructed SFHs represent trajectories in SFR-M$_*$ space, enabling us to study galaxies at epochs earlier than observed by propagating them backwards in time along these trajectories. We study the SFR-M$_*$ correlation at $z=1,2,3,4,5,6$ using both direct fits to galaxies observed at those epochs and SFR-M$_*$ trajectories of galaxies observed at lower redshifts. The SFR-M$_*$ correlations obtained using the two approaches are found to be consistent with each other through a KS test.  Validation tests using SFHs from semi-analytic models and cosmological hydrodynamical simulations confirm the sensitivity of the method to changes in the slope, normalization and shape of the SFR-M$_*$ correlation. This technique allows us to further probe the low-mass regime of the correlation at high-z by $\sim 1$ dex and over an effective volume of $\sim 10\times$ larger than possible with just direct fits. We find that the SFR-M$_*$ correlation is consistent with being linear down to M$_*\sim 10^7 M_\odot$ at $z>4$. The evolution of the correlation is well described by $\log SFR= (0.80\pm 0.029 - 0.017\pm 0.010\times t_{univ})\log M_* - (6.487\pm 0.282-0.039\pm 0.008\times t_{univ})$, where $t_{univ}$ is the age of the universe in Gyr.

\end{abstract}




\section{Introduction} \label{sec:intro}

The SFR-M$_*$ correlation couples a galaxy's Star Formation Rate (SFR), an effectively instantaneous quantity, to its stellar mass (M$_*$), accumulated over its lifetime \citep{noeske2007star, daddi2007multiwavelength, elbaz2007reversal, salim2007uv}. The correlation persists across a wide range of stellar masses and SFRs and over a range of redshifts \citep{whitaker2012star, speagle2014highly, salmon2015relation, tasca2015evolving, schreiber2015herschel, kurczynski2016evolution, johnston2015evolving, santini2017star}. This has led to speculations about its origin, with theories suggesting this is controlled by halo mass accretion \citep{dutton2010origin, forbes2014origin, rodriguez2015main}, or the regulation of gas infall and feedback \citep{tacchella2016confinement, mitra2016equilibrium}, or that the observed correlation is simply a cross-section of a more fundamental SFR-M$_*$-Z relation, with the evolution explained by increasing metallicities with cosmic time \citep{mannucci2010fundamental, lilly2013gas, torrey2017similar}.

Galaxy formation models predict the evolution of individual star-forming galaxies comprising the SFR-M$_*$ correlation. Testing these predictions has been difficult, with observations up to now unable to reveal if individual galaxies evolve along the correlation, as assumed by the `Main Sequence Integration' technique \citep{leitnerMSI, munoz2015framework} or make significant excursions above and below it \citep{pacifici, tacchella2016confinement}.

One of the most common ways of probing the SFR-M$_*$ relation is estimating the stellar mass and SFR through SED fitting, with modern techniques capable of handling large quantities of data and extracting high-fidelity information. A key improvement in obtaining the stellar masses and SFRs comes from relaxing the assumption that a galaxy's Star Formation History (SFH) be described by a single parametric form such as exponentially declining or constant star formation \citep{iyer2017reconstruction, pacifici, acquaviva2011sed, lee2017intrinsic, ciesla2017sfr}. Although the stellar masses and SFRs obtained through SED fitting offer a probe of the SFR-M$_*$ correlation to extremely high redshifts, it is sensitive to the systematic assumptions inherent in SED fitting, as well as the decreasing S/N as we approach dimmer objects and higher redshifts. To probe the correlation in this regime, we thus need to probe beyond these traditionally estimated SED-fit quantities.

In this paper, we apply the Dense Basis SED fitting method developed in \citet{iyer2017reconstruction}, which reconstructs a galaxy's Star Formation History using the best-fit from among multiple families of smooth SFHs. We use these SFHs to construct SFR-M$_*$ trajectories, along which individual galaxies are propagated backwards in time to reveal the SFR-M$_*$ diagram at higher redshifts. 
This method allows us to gain statistical power from the large number of low-redshift galaxies that can contribute to the SFR-M$_*$ correlation at higher redshifts. In contrast to direct fits, these trajectories possess the advantage of probing lower masses as we extend the method to higher redshifts. Currently, the lowest stellar masses the correlation has been probed at involve using galaxies observed in the Hubble Ultra Deep Field to go down to $10^7M_\odot$ at $z \sim 1.5$ \citep{kurczynski2016evolution}, or using gravitationally lensed galaxies in the HST Frontier Fields to go down to $10^{8.8} M_\odot$ at $z\sim 6$ \citep{santini2017star}. In this paper, we show that our technique of reconstructing SFR-M$_*$ trajectories allows us to recover the SFR-M$_*$ correlation down to $10^6 M_\odot$ at $1<z<6$.

The paper is structured as follows; in $\S $\ref{sec:data}, we specify the choice of dataset, followed by the details of our analysis in $\S $\ref{sec:method}. In $\S $\ref{sec:results} we present our main results, describe the validation tests we performed in $\S $\ref{sec:validation}, and discuss the implications and caveats in $\S $\ref{sec:systematics}. Throughout this paper magnitudes are in the AB system; we use a standard $\Lambda$CDM cosmology, with $\Omega_m = 0.3$, $\Omega_\Lambda = 0.7$ and $H_0 = 70$ km Mpc$^{-1}$ s$^{-1}$.

\section{Dataset} \label{sec:data}

We fit 17-band photometric SEDs\footnote{The photometric bands used are: U(CTIO), U(VIMOS), HST/ACS F435w, F606w, F775w, F814w, F850lp, HST/WFC3 F098w, F105w, F125w, F160w, VLT/HAWK-I Ks, VLT/Isaac Ks, and Spitzer/IRAC 3.6, 4.5, 5.8, 8$\mu$m. \citep{guo2013candels} } spanning the rest-frame UV through near-IR (IRAC) from  the Cosmic Assembly Near-Infrared Deep Extragalactic Legacy Survey (CANDELS; \citet{candels, koekemoer2011candels, nonino2009deep, retzlaff2010great, fontana2014hawk, ashby2015s}). \added{The catalog selects objects in the GOODS-S field via SExtractor in dual-image mode using F160w as the detection band. The dual image mode \citep{galametz2013candels} is optimized to detect both faint, small galaxies in `hot' mode without over de-blending large, resolved galaxies in `cold' mode. The HST (ACS and WFC3) bands were point spread function (PSF) matched to measure photometry, and TFIT \citep{laidler2007tfit} was used to measure the photometry of ground based and IRAC bands using the HST WFC3 imaging as a template.} We consider galaxies in the GOODS-S field \citep{guo2013candels} at redshifts $0.5<z<6$. 
The three different depths (wide, deep, HUDF) allow us to probe a population of galaxies across a wide range of masses to construct SFR-M$_*$ trajectories. 
To reduce the effects of incompleteness for the sample while accounting for the different depths, we require our sample to be brighter than a F160w magnitude limit of $25.9,~ 26.6$ and  $28.1$ in the wide, deep and HUDF regions, respectively \citep{guo2013candels}.

\begin{figure}
    \centering
    \includegraphics[width=220px]{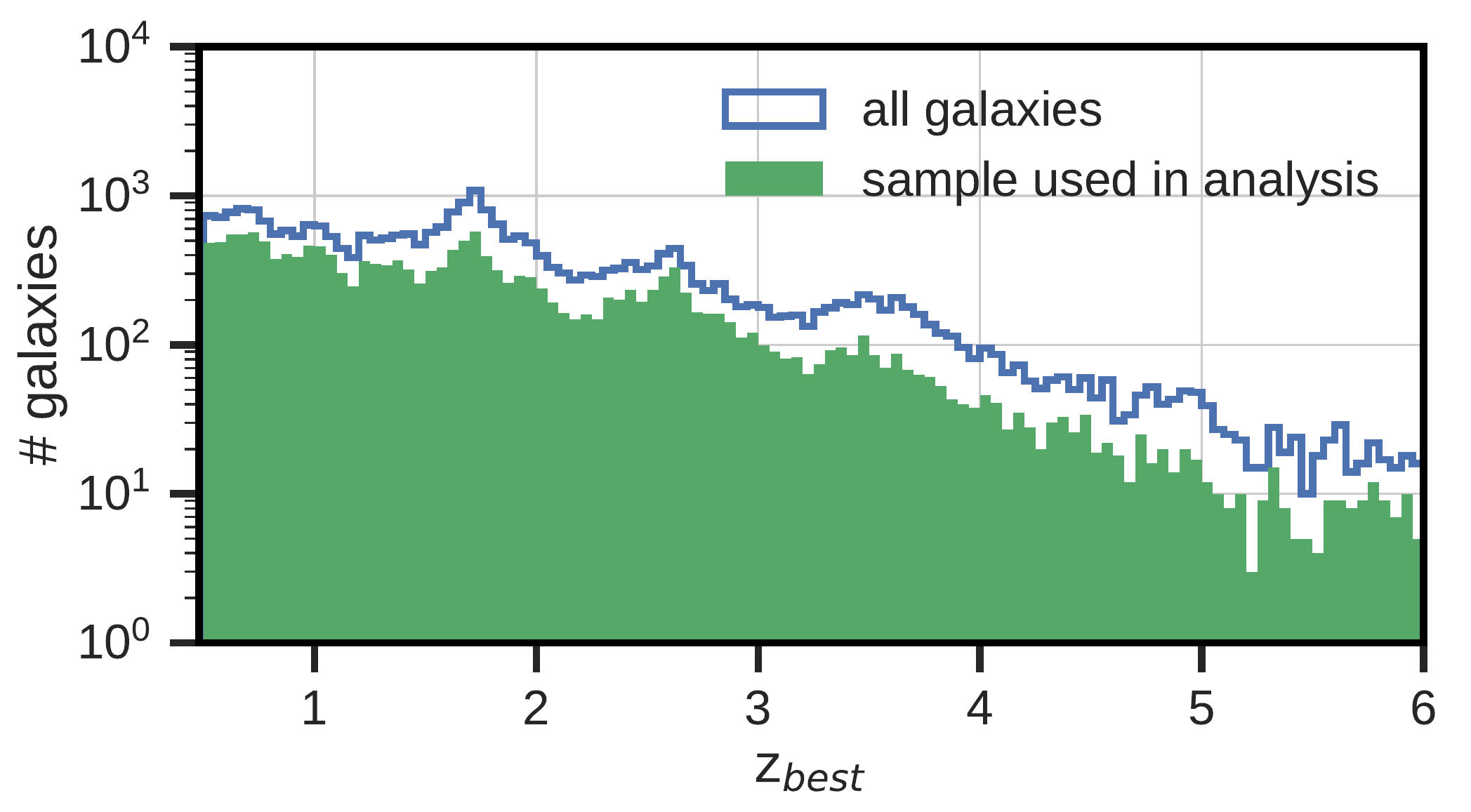}
    \caption{Redshift distribution of the CANDELS/GOODS-S galaxies at $0.5<z<6.0$ (blue line) and for the sample we use in our analysis (solid green histogram) excluding bad fits and galaxies with F160w magnitudes $<$ $25.9,~ 26.6$ and  $28.1$ in the wide, deep and HUDF regions, respectively.}
    \label{fig:zbest_dist}
\end{figure}

In performing our fits, we exclude objects that are marked as stars or X-ray detected AGN \citep{hsu2014candels}
in the \citet{santini2015stellar} mass catalog \added{(487 objects)}, as well as poor fits in our SED fitting routine ($\chi^2 > 50$, \added{595 objects}). After these cuts, our analysis includes $\sim 17,873$ galaxies \added{(94.7\% of our parent sample)}.  
To perform our fits, we use an updated photometric redshift catalog by Kodra et al (in prep.) containing an increased number of spec-z measurements as well as photometric redshifts with Bayesian combined uncertainties estimated by comparing the redshift probability distributions of four different SED fitting methods. Figure \ref{fig:zbest_dist} shows the redshift distribution of our analysis sample.
We perform our fits using their $z_{\rm best}$\deleted{footnote: insert explanation for what z$_{\rm best}$ is - hierarchical Bayesian combination of photo-z pdfs from different methods, and spec-z or grism-z where available.} binned to the resolution of our pre-grid, with $\delta z = 0.01$. The $z_{\rm best}$ in our redshift range includes $\sim 1917$ spectroscopic and $\sim 384$ grism redshifts, in addition to photometric redshifts. \deleted{Since the majority of galaxies in our sample have photometric redshifts, we adopt redshift bins of width $\Delta z = 0.3$ for our analysis.} 

\section{Methodology} \label{sec:method}

\subsection{SED fitting}

\begin{figure*}[ht!]

\begin{center}
\includegraphics[width=250px]{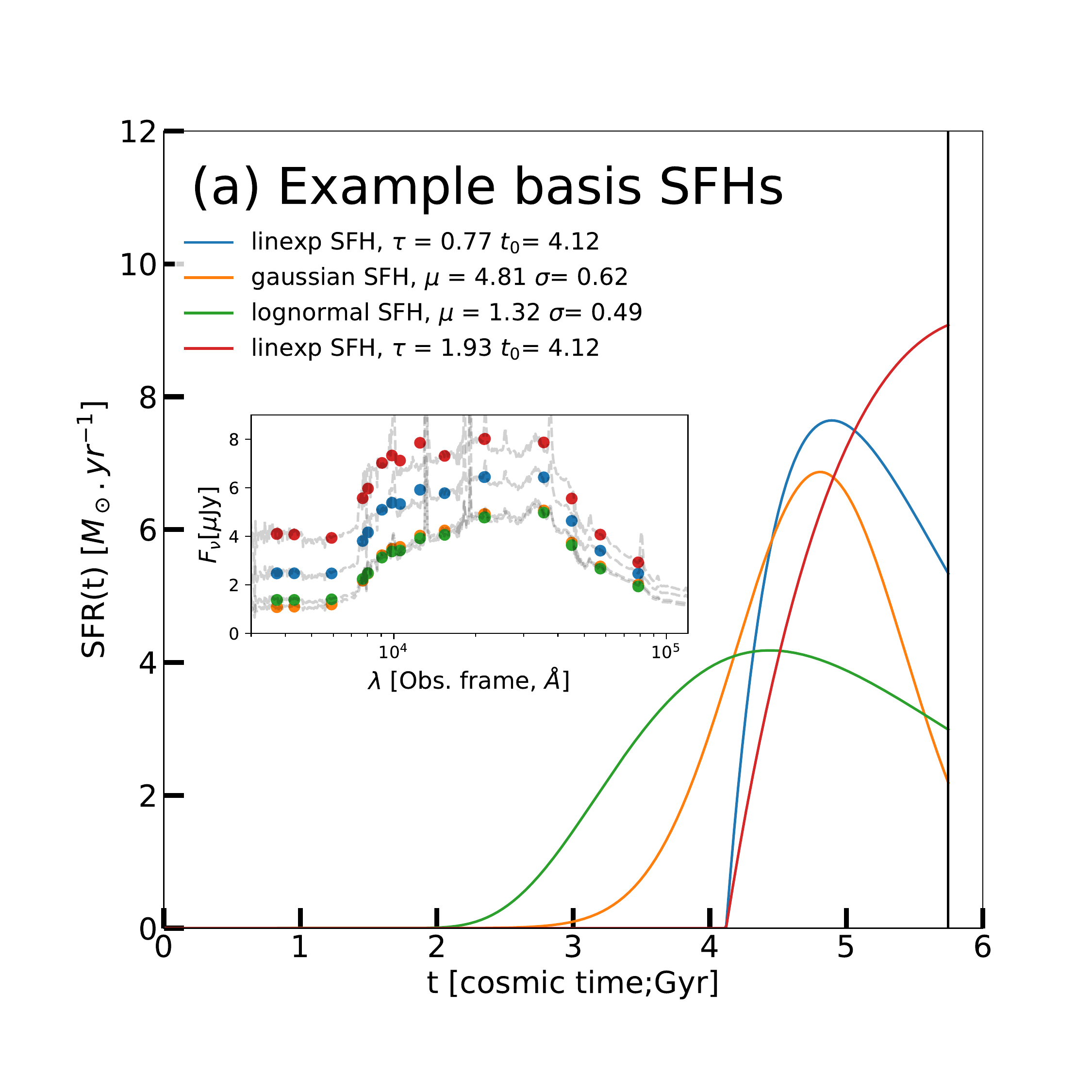}
\includegraphics[width=250px]{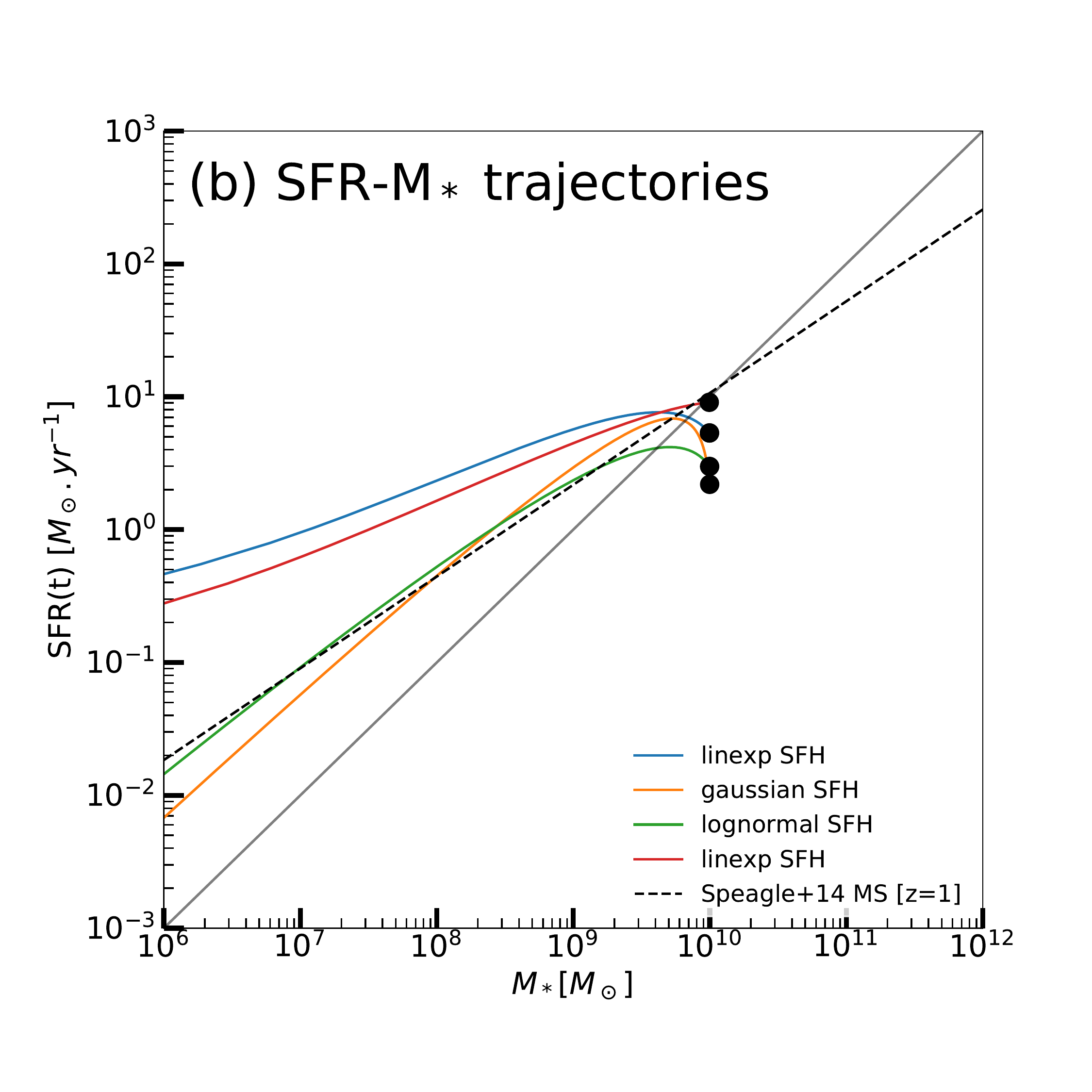}
\includegraphics[width=250px]{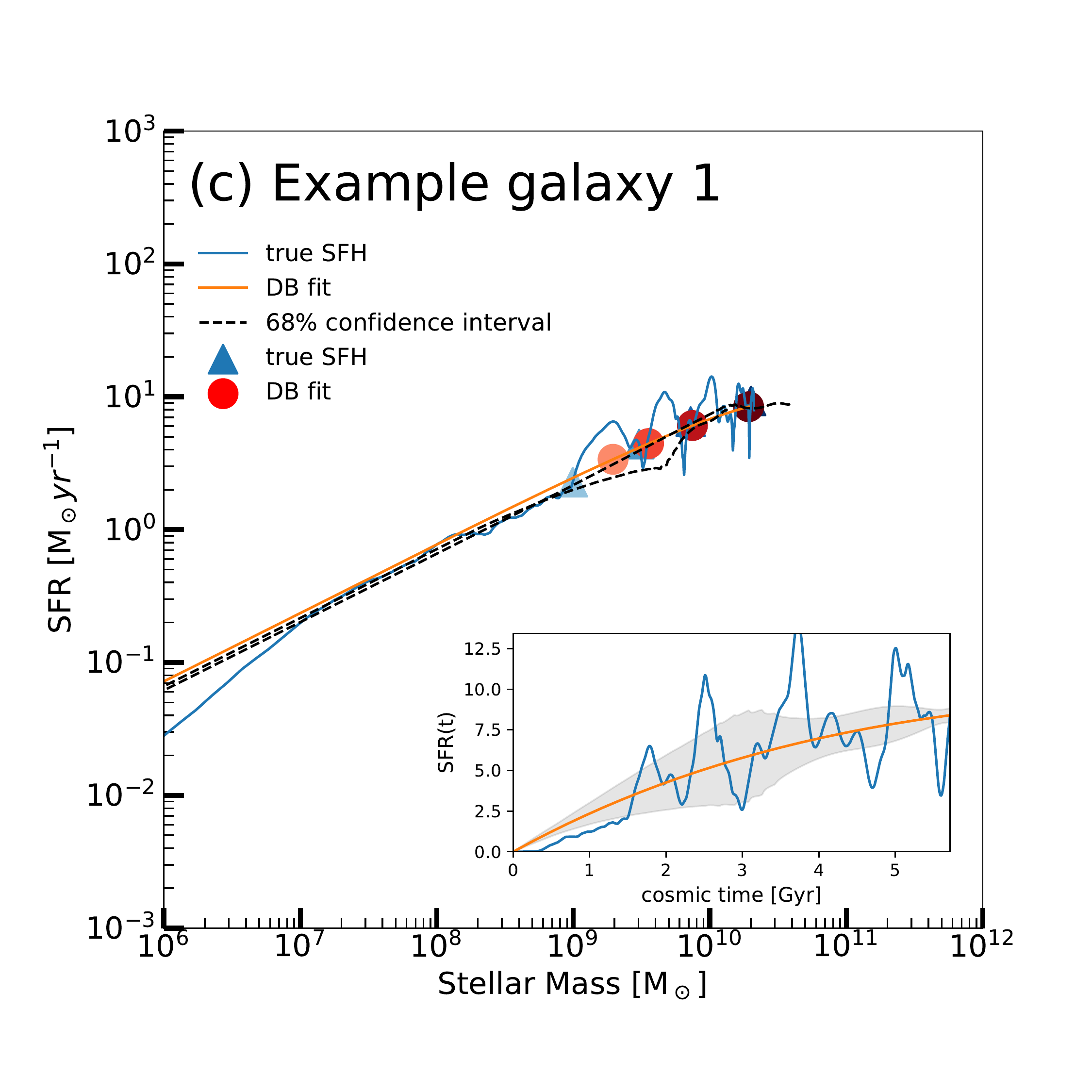}
\includegraphics[width=250px]{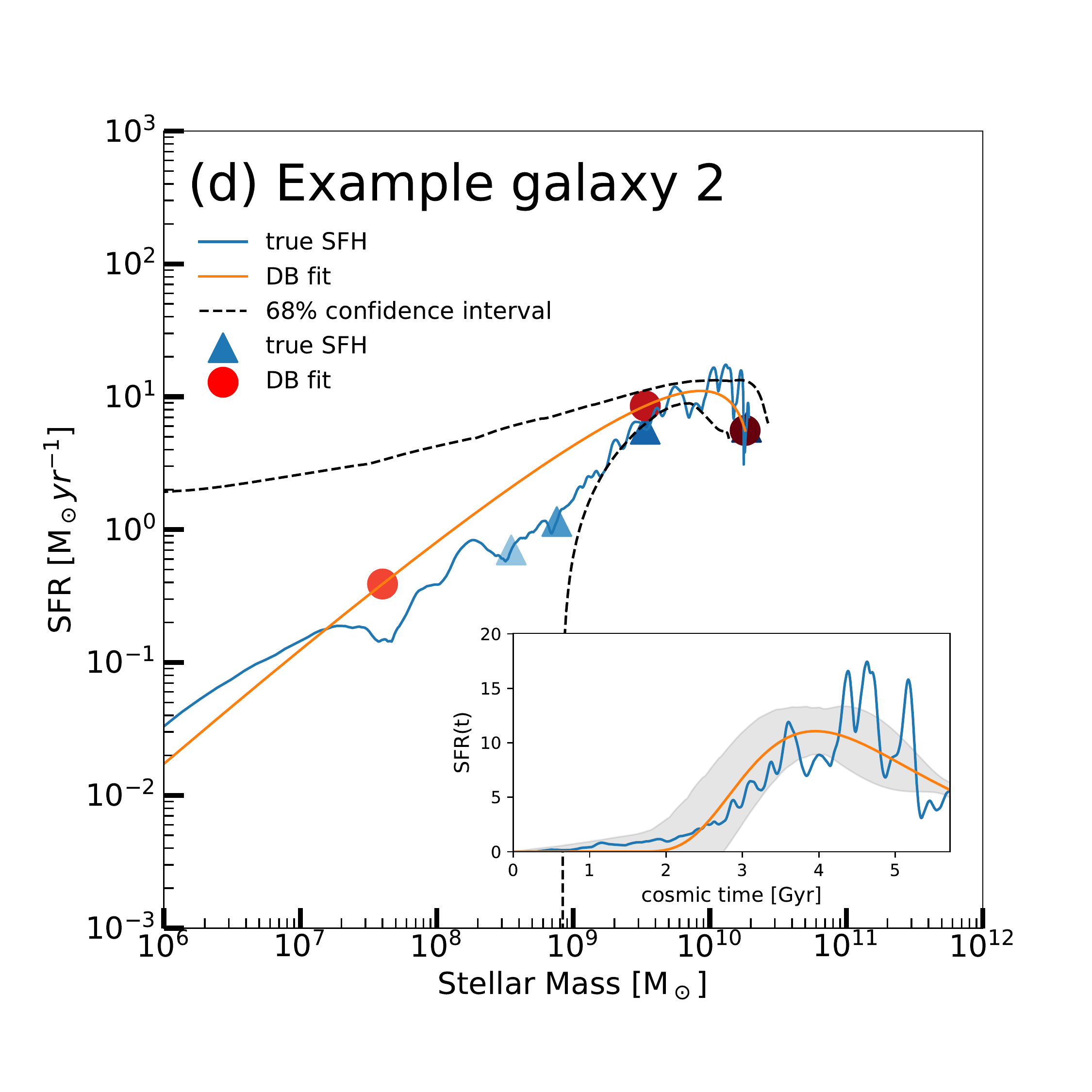}
\caption{\textbf{(a)} Examples of basis Star Formation Histories belonging to the three functional families used for SED fitting in the paper, normalized to the same stellar mass of $10^{10}M_\odot$, with similar SFRs. Inset panel shows corresponding SEDs observed at z=1\added{ with reference spectra plotted vs observed wavelength}. \textbf{(b)} Trajectories in the SFR-M$_*$ plane corresponding to the \replaced{three}{four} SFHs shown above, with black circles illustrating their locations when lookback time equals zero. The dashed black line shows the \citet{speagle2014highly} SFR-M$_*$ relation at z=1 \added{and the solid grey line shows $\log SFR = \log [M_*/10^9M_\odot]$ for reference}. \textbf{(c,d)} Examples of SFH reconstructions of individual $z=1$ MUFASA (panel c) and SAM (panel d) galaxies with their uncertainties, extended to trajectories in SFR-M$_*$ space. Inset figures show the simulated SFH (true SFH, blue) and the SFH reconstructed through SED fitting the noisy simulated photometry (DB fit, orange). The main panels show their corresponding trajectories in SFR-M$_*$ space. Coloured circles and triangles show SFR, M$_*$ estimates at z=1,2,3,4 (darker to lighter colors), and dashed lines show the 68\% confidence interval, corresponding to the grey shaded region in the inset. \added{Example galaxy 2 (panel d) can be reliably propagated back to $z\sim 2$, but not beyond that. In our analysis, we exclude such trajectories at redshifts where they have large uncertainties.}
}
\label{fig:basis_sfhs}
\end{center}
\end{figure*}

The Dense Basis method described in \citet{iyer2017reconstruction} uses a physically motivated basis of Star Formation Histories to generate an atlas of template SEDs. The best-fit SFH, dust and metallicity values for each observed galaxy SED are computed using standard $\chi^2$ minimization over the entire atlas. To generate spectra corresponding to a galaxy with a given basis SFH, we use the Flexible Stellar Population Synthesis (FSPS) model \citep{conroy2009propagation, conroy2010propagation, dan_foreman_mackey_2014_12157}.
We use a Chabrier IMF \citep{chabrier2003galactic}, Calzetti dust reddening \citep{calzetti2001dust}, and IGM absorption according to the \citet{madau1996high} prescription. 
Star Formation Histories are drawn from the Linexp (linear rise followed by exponential decline, sometimes called Delayed-$\tau$ models), Gaussian and Lognormal families of curves, excluding SFHs that are extremely similar in shape to reduce the size of the basis, with a slight modification of \citet{iyer2017reconstruction}, where we also considered Bessel function rise followed by exponential decline (Bessel-exp), Top-hat (Exp) and Constant Star Formation (CSF) histories. In this work, we do not consider these three families because Bessel-exp SFHs are extremely similar in shape to the SFHs we already consider in our basis, and Exponential and CSF were shown in \citet{iyer2017reconstruction} to lead to biased estimates of galaxy properties. The parameter ranges for the various families are given in Table.\ref{table:param_range}. \added{Defining the slope as a tangent to the log SFR-log M$_*$ trajectory at a given point,} the range of SFH shapes lead to trajectories that can have a wide range of slopes  at low masses $\sim [0,16]$ as well as flat and negative slopes $\sim [-34,52] $ at high masses, as galaxies enter a quiescent phase. Figure.\ref{fig:basis_sfhs} (a,b) shows examples of SFHs from each of the three families, as well as their corresponding trajectories in SFR-M$_*$ space. Insets in panels (c,d) of the same figure show examples of mock SFHs from the MUFASA hydrodynamic simulation (panel c, \citet{dave2016mufasa}) and a Semi-Analytic Model (panel d, \citet{somerville2015star}) and their reconstructed best-fit SFHs with uncertainties. The mock SEDs fitted to reconstruct the SFHs were generated using the same filters as the CANDELS/GOODS-S catalog, with realistic photometric noise and dust, as in \citet{iyer2017reconstruction}. While the method does not recover short stochastic episodes of star formation, it does well approximate the overall trend of the galaxy's SFH, thus allowing us to construct robust trajectories in SFR-M$_*$ space. Since we fit galaxies with a single episode of star formation, we find that our reconstructions \added{can sometimes} fail in cases where the true SFH of the galaxy contains multiple strong episodes of star formation. 
\added{However, in} \citet{iyer2017reconstruction} we find that only about 15\% of galaxies at z$\sim$1 support fits with two episodes of star formation.
\deleted{Although this is capable of causing a small change in the slope and normalization estimated using our method, we expect the effect to become weaker as we go to higher redshifts.}
Fits to the galaxy SEDs at different redshifts provide us with both the Stellar Masses and effectively instantaneous Star Formation Rates at the epoch of observation, referred to as `Direct Fits' for the rest of this work. In addition to this, the reconstructed Star Formation Histories are then used to construct SFR-M$_*$ trajectories, which allow us to infer the Stellar Masses and Star Formation Rates at higher redshifts of interest.

\begin{table}[ht!]
\caption{Parameter ranges for SFH families\footnote{Parameter ranges, in Gyr, adapted for $0.5<z<6$ from \citet{iyer2017reconstruction}. In addition to these, there is a normalization corresponding to the stellar mass, which can be considered a third free parameter in specifying the SFH.}}
\label{table:param_range}
\begin{center}
\begin{tabular}{ c|c c }
\hline \hline
SFH &	param 1  &	param 2   \\
\hline
Linexp		&	$\tau \in [0.05,10]$	&	$t_0 \in [0,t_{univ}]$ \\
Gaussian		&	$\mu \in [0,t_{univ}]$ & $\sigma \in [0.1,10]$ \\
Lognormal	&	$\mu \in [0,t_{univ}]$ & $\sigma \in [0.1,2]$ \\
\hline
\end{tabular}
\end{center}
\end{table}

\begin{figure*}[ht!]

\begin{center}
\includegraphics[width=240px]{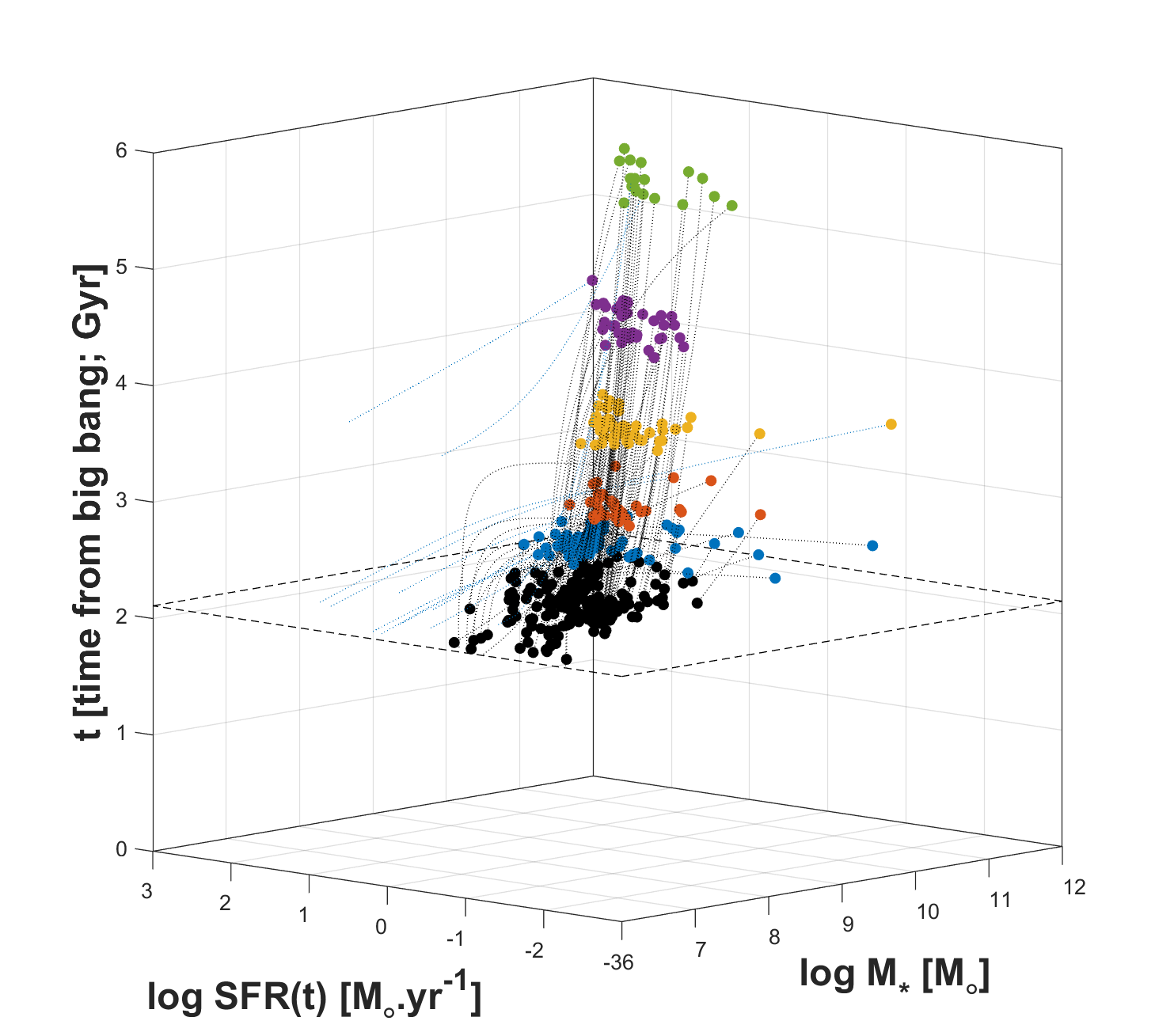}
\includegraphics[width=240px]{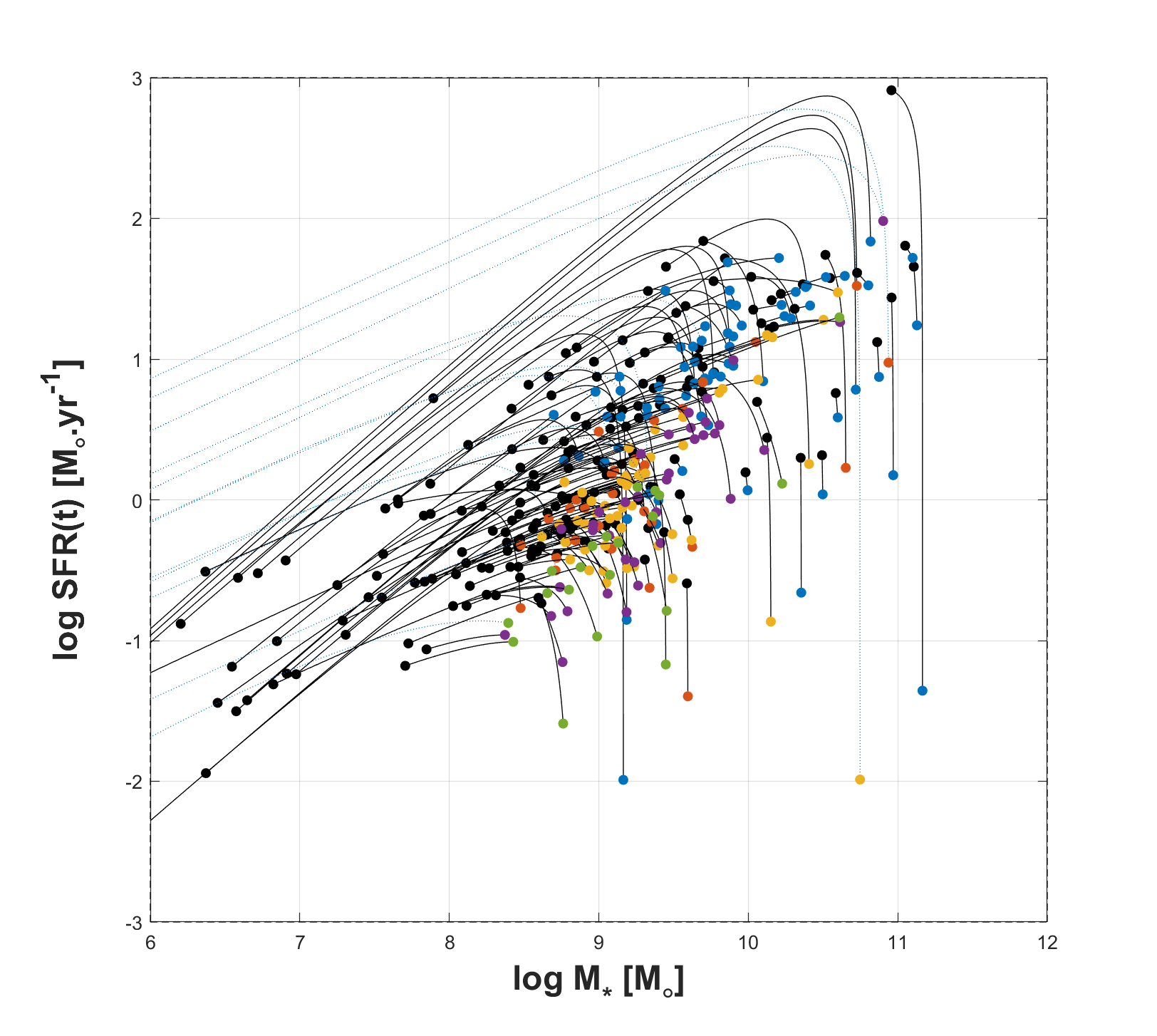}
\includegraphics[width=420px]{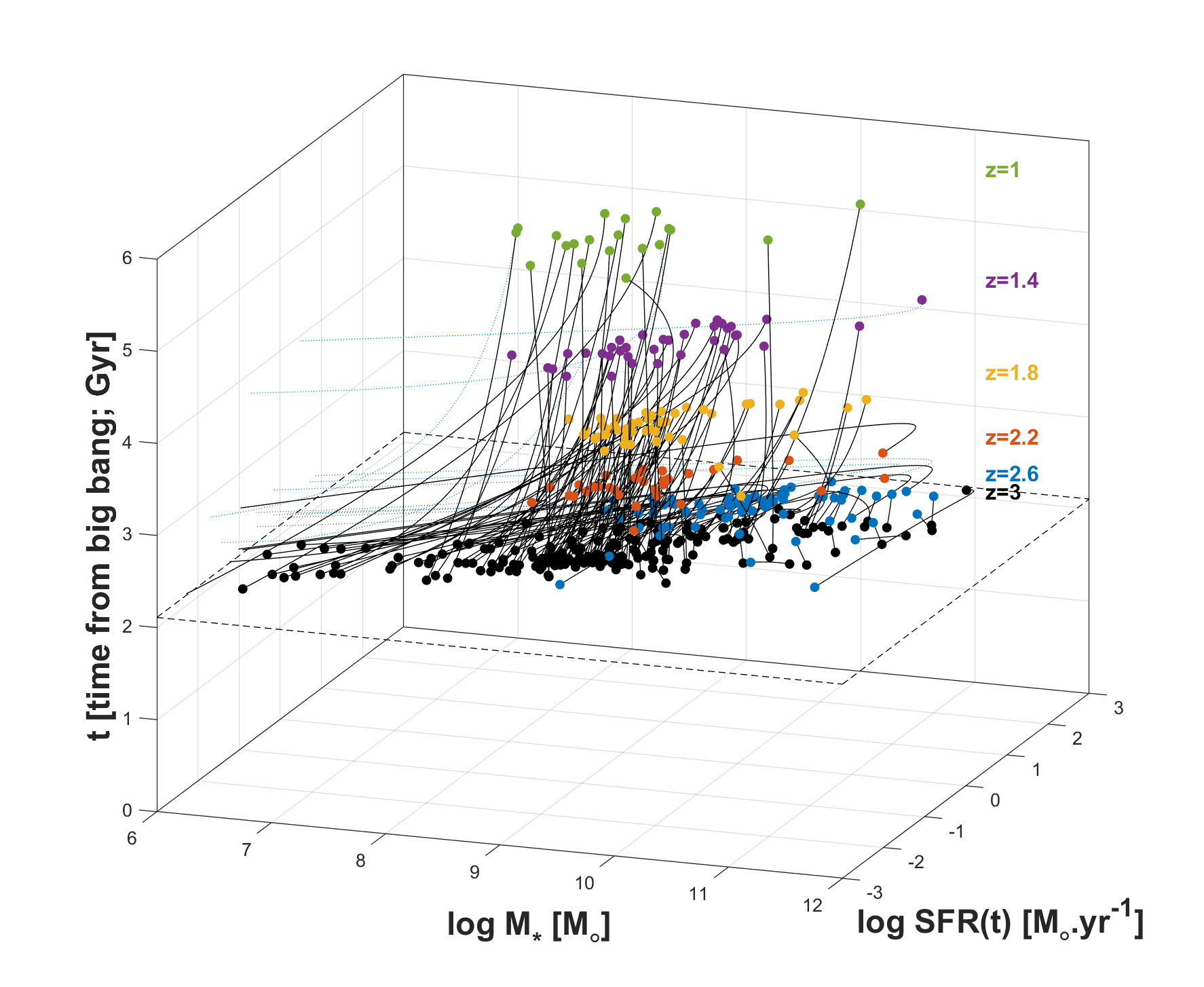}

\caption{The SFR-M$_*$ correlation at $z=3$ (black points) constructed by propagating a randomly chosen subset of galaxies at redshifts $z = [1.0,1.4,1.8,2.2,2.6]$ (colored points) backwards in time along their best-fit SFR-M$_*$ trajectories \replaced{(black dotted lines)}{(black lines are galaxies whose trajectories go back to z=3, blue dotted lines are galaxies whose trajectories drop off the plot at lower redshifts). Trajectories that go beyond $z=3$ are truncated at $z=3$ for clarity.}  Three orientations of the figure are shown along different viewing angles. The plots allow us to see that the correlation evolves along a relatively narrow phase space, although individual galaxies can make significant excursions above and below it. This allows us to probe the correlation at high redshifts using higher S/N SED fits at lower redshifts. It also allows us to probe the correlation down to lower masses than previously possible. 
}
\label{fig:sfr_mstar_3dplot}
\end{center}
\end{figure*}

\subsection{SFH uncertainties:}

We compute uncertainties on the reconstructed SFH using the full $\chi^2$ surface over the multidimensional space spanning SFH, dust and metallicity, generated through the fitting procedure, following a method similar to \citet{iyer2017reconstruction}. Using the cumulative histogram of the $\chi^2$ values, we select the 100 SFHs corresponding to the set of lowest $\chi^2$ values among the set and compute the median, which we hereafter refer to as the median SFH. We find that changing this threshold of 100 SFHs does not substantially affect the computed uncertainties for this analysis. \replaced{We then prune this set of SFHs,}{We then prune the set of good SFHs,} removing those that are simply bad fits ($\chi^2 > 10*min(\chi^2)$) or outliers in SFH space $(max(|SFH_i - SFH_{median}|)/\langle SFH_{median} \rangle > 5)$, similar to any robust algorithm that is insensitive to outliers. We use the distribution of remaining SFHs to derive pointwise 68\% confidence intervals in time for the reconstructed SFH, as well as a robust median SFH. We use the stellar masses and SFRs of the full set of good fits to derive confidence intervals for those quantities at any lookback time along the trajectory. \added{ This technique is also used to find the uncertainties on SFR and Stellar Mass at lookback times corresponding to the redshifts of interest in Sec. \ref{sec:results}. To check that our uncertainties are robust at all lookback times as a function of rest-frame wavelength coverage (depending on $z_{obs}$) or SED S/N, we use mock SEDs to analyze possible biases in estimating SFR and M$_*$ as a function of lookback time for different $z_{obs}$. This is detailed further in Appendix \ref{app:validation_contd}, where we find that our median uncertainties are conservative and increase accordingly at lookback times where the Stellar Mass or SFR is poorly constrained.}

\subsection{SFR-M$_*$ trajectories:}

The best-fit reconstructed Star Formation Histories are curves of SFR(t) against time. At any instant in time, the Stellar Mass is given by 
\begin{equation}
M_*(t) = \int_0^{t} SFR(t') f_{ret}(t'-t,Z) dt
\end{equation}
where $f_{ret}(t'-t,Z)$ is a metallicity dependent fraction of the mass of formed stars that is retained as stars or stellar remnants at the time of observation obtained from FSPS \citep{conroy2009propagation}, which is typically between 0.6-1.0. Using this relation and the best-fit SFH, we can construct a parametrized curve corresponding to [SFR(t),M$_*$(t),t], which provides a trajectory in SFR-M$_*$ space. Observing this trajectory at any redshift gives the [SFR(z),M$_*$(z)], allowing us to extend trajectories to higher redshifts and fill in the SFR-M$_*$ correlation using previously inaccessible data from earlier periods in a galaxy's lifetime. Panel (b) of Figure \ref{fig:basis_sfhs} shows examples of trajectories corresponding to each of the basis SFHs shown in panel (a). Panels (c,d) of Figure.\ref{fig:basis_sfhs} show \added{a couple of} examples of reconstructed SFR-M$_*$ trajectories corresponding to a couple of SFHs at $z=1$ from a hydrodynamical simulation (MUFASA, \citet{dave2016mufasa}) and a Semi-Analytic Model \citep{somerville2015star, somerville2008semi}. As seen in the figure, the true trajectory is well approximated by the smooth reconstruction and its corresponding uncertainties, matching the observations not only at the epoch of observation ($z=1$), but also at earlier epochs ($z=2,3,4$). \added{While the galaxy in panel (c) has a SFH that can be traced back to very high redshifts, this is not in general true for most observed galaxies. Galaxy trajectories may fail to contribute meaningfully at higher redshifts either because they drop off the plot, ie. they formed most of their mass at more recent epochs, or because their uncertainties grow extremely large. The latter case is illustrated through the example galaxy in panel (d), which can be reliably traced back to $z\sim 2$, but has large uncertainties beyond that. In our analysis, we exclude such trajectories at redshifts where they have large uncertainties.}

Figure.\ref{fig:sfr_mstar_3dplot} shows a randomly selected sample galaxies at $1<z<3$, that are propagated backwards in time along their trajectories to infer the SFR-M$_*$ correlation at $z\sim 3$. \added{Black lines denote trajectories for galaxies that can be propagated backwards to $z=3$, while blue dotted lines denote trajectories for galaxies whose trajectories do not reach back to $z=3$. Since we choose a random subsample of galaxies to plot at each redshift, the F160w selection threshold results in the appearance of a rising lower limit in stellar mass as we go to higher redshifts - this doesn't imply that galaxies are more massive at $z \sim 2.6$, but that observationally selected galaxies, of which we pick a random sample, tend to be the more massive ones. The average amount of time a galaxy is propagated backwards in time along its trajectories shows a mild increase as we go to higher redshifts but remains much smaller than the amount of time between $z=0.5$ and the redshift of interest, as shown in appendix \ref{app:validation_contd}.}

\added{For the rest of this work, while considering a sample of galaxies propagated backwards in time along their SFR-M$_*$ trajectories, we restrict ourselves to the sub-sample of galaxies with low uncertainties ($\sqrt{\sigma_{SFR}^2 + \sigma_{M_*}^2} < 1$ dex) to minimize the effects of possible biases. This is explored in detail in Appendix \ref{app:validation_contd}, where we fit mock SEDs corresponding to SFHs from simulations to assess the robustness of SFR and Stellar Mass as we propagate galaxies backwards in time along their trajectories. In doing so, we find that the uncertainties closely trace possible biases, incorporating effects due to factors like S/N and the rest-frame wavelength coverage during SED fitting. This allows us to isolate a subsample with minimal bias that we use for the analysis in this paper.}

\section{Results: The SFR-M$_*$ correlation from direct fits and trajectories} \label{sec:results}

\begin{figure*}[ht!]

\begin{center}

\includegraphics[width=420px]{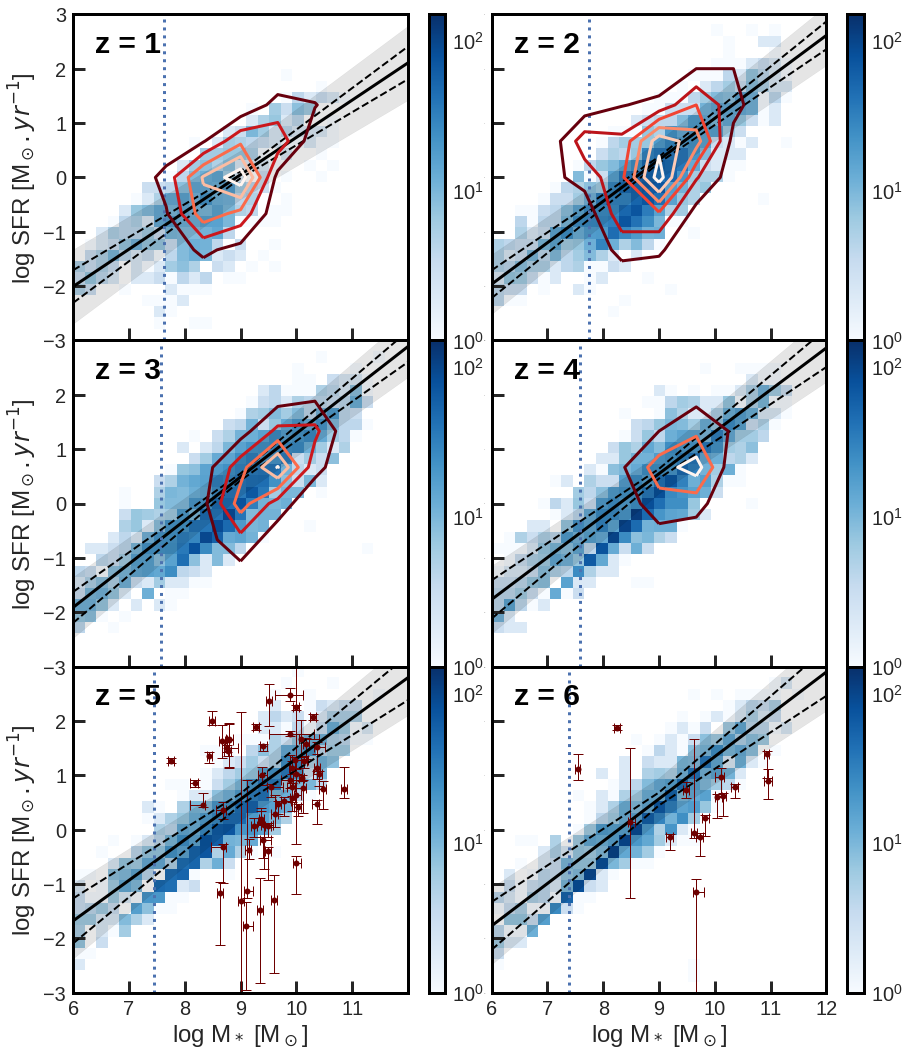}
\caption{The SFR-M$_*$ correlation at $z = [1,2,3,4,5,6]$. Galaxies observed at the epochs of interest (direct fits) are shown as \replaced{orange}{red} contours for $z<5$, and as \replaced{orange}{red} points with error bars for $z=5,6$ where there are insufficient points to yield representative contours. Galaxies observed at later epochs and propagated backwards in time along their SFR-M$_*$ trajectories are shown as the colored heatmap, with the colorbars denoting the number of galaxies in a particular pixel. \deleted{To show the overall trend of the relation, points are binned along the $\log SFR = \log [M_*/10^9 M_\odot]$ line, with the median and 68\% interval in each dimension shown for each bin with more than 20 objects for direct fits (orange errorbars), trajectories (blue errorbars; shifted by 0.2 dex for clarity) and the combined dataset (white errorbars).} The \replaced{white}{black} solid line shows our best-fit to the combined dataset, with uncertainties \added{denoted by the dashed black lines.}. The shaded black region shows the uncertainties + observed scatter around the best-fit. \added{Dotted blue lines show the 10$^{th}$ percentile in stellar mass for trajectories.} \added{Additional non-parametric fits to the correlation are shown in Appendix \ref{app:nonparametric_fits_sfr_mstar}.} We see that the SFR-M$_*$ correlation is consistent with being linear out to very low masses and high redshifts.
}
\label{fig:mainseq_allz}
\end{center}
\end{figure*}

In Figure \ref{fig:mainseq_allz}, we present the SFR-M$_*$ correlation at z = [1,2,3,4,5,6], including estimates from galaxies observed at those epochs (henceforth direct fits), as well as from galaxies observed at later epochs propagated backwards in time along their trajectories (henceforth trajectories). The direct fits are shown as contours and as individual datapoints at high redshifts where the number of galaxies are small. The contribution from trajectories at each redshift is shown as a coloured heatmap. 
\deleted{To determine the general trend of the correlation, we bin both direct fits and trajectories in bins of 0.5 dex perpendicular to the diagonal $\log SFR = \log [M_*/10^9M_\odot]$ line, and show the median and 68\% intervals for each bin, for bins that contain more than 20 galaxies. We adopt this choice of binning to minimize the influence of selection effects in estimating the slope of the correlation. 
These are shown as orange points for direct fits, blue points for trajectories, and white points for the combined dataset of direct fits and trajectories.}

We find that the locus of the direct fits and trajectories broadly agree with each other. \replaced{A KS test of the distribution of distances of individual galaxies from the combined best-fit SFR-M$_*$ correlation for both directs fits and trajectories at each redshift shows that the difference between the two distributions is not statistically significant}{To quantify this statistically, we perform a KS test comparing the two datasets at each redshift of interest}, as shown in Table \ref{table:dist_compare}. Since we do not want to compare outlier distributions, due to starbursts or quenched galaxies, we impose a cutoff, excluding galaxies that are at a distance $\geq 0.4$ dex from the best-fit SFR-M$_*$ correlation. \added{In Appendix \ref{app:kstest_diffsamples} we also compare distributions using a variable threshold based on the observed scatter and find that the results do not change.} \replaced{The fraction of total galaxies at different epochs as well as the fraction of galaxies}{The number of galaxies} that contribute to the SFR-M$_*$ correlation \added{from direct fits and trajectories} are given in the table. \added{The p-values for all these comparisons are $> \alpha $, indicating that the two distributions are not statistically different. The
significance level for each test is $\alpha' = (0.05 / 6) \approx 0.0083$, where we apply a Bonferroni correction  \citep{goeman2014multiple} to control for false positives since we are performing a family of tests to evaluate a single hypothesis. Since our results remain consistent across this broad range of tests, we can not reject the null hypothesis that the two samples are drawn from a common underlying distribution at $> 95\%$ confidence. While this does not completely rule out the possibility that the two distributions are different,} this agreement justifies the usage of a combined sample to obtain our primary results. \added{We further explore the comparison between the two distributions in Appendix \ref{app:kstest_diffsamples}.}

We plot the best-fit line to \replaced{this combined dataset}{the combined dataset of direct fits and trajectories} in Figure \ref{fig:mainseq_allz}, determined using an iterative robust fitting routine that excludes outliers \citep{holland1977robust}. \added{To compare trajectories and direct fits on the same footing, the direct fits are analyzed at z=[1,2,3,4,5,6] in bins of $\Delta z = 0.1$.} The uncertainties are determined using 1000 Monte Carlo realizations of the data perturbed within the M$_*$ and SFR error estimates obtained through the Dense Basis SED fitting routine. We find that the SFR-M$_*$ correlation extends to redshifts as high as $z\simeq 6$ \citep{steinhardt2014star}, and remains linear down to masses as low as $\log M_*/M_\odot \sim \replaced{6}{7}$, which is a factor of \replaced{100}{10} below current estimates from direct fits. To test the linearity of the correlation, we fit the combined data at each redshift with polynomials of order 1 (linear) and 2 (quadratic) and see if the corresponding improvement in the goodness-of-fit is statistically significant using an F-test. At all redshifts, we find that the linear fit is preferred, at $> 90\%$ confidence, with values at individual redshifts given in Table \ref{table:slope_norm_evol_table}. From the reconstructed SFHs, $\sim 92\%$ of the galaxies in the sample have $t_{10} \leq 3 Gyr$ and $\sim 70\%$ of the galaxies in the sample have $age \leq 3 Gyr$, where $t_{10}$ is the lookback time at which they formed the first 10\% of their observed stellar mass. This implies that most of the galaxies that contribute to the SFR-M$_*$ diagram form the majority of their stellar mass in $\leq 3$ Gyr in the redshift range we consider, with $70\%$ entering the observable SFR-M$_*$ range within that time. \added{To account for the fact that the F-test need not guarantee that the correlation is indeed linear, we also perform non-parametric regression in Appendix \ref{app:nonparametric_fits_sfr_mstar}. Using this, we see that the nonparametric methods closely approximate the best-fit line as we go to low stellar masses at high redshifts.}

\begin{figure*}[ht!]

\begin{center}
\includegraphics[width=166px]{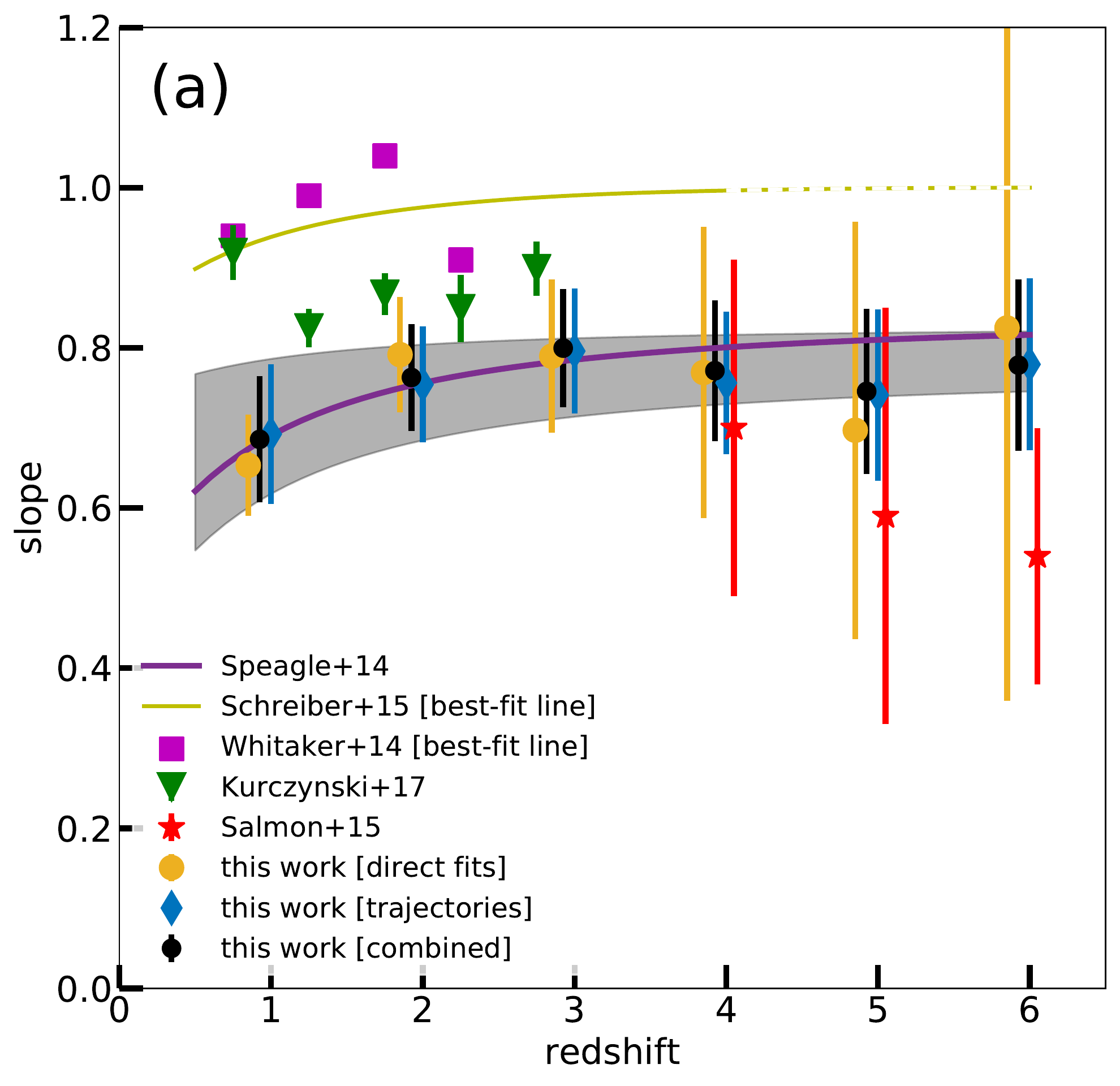}
\includegraphics[width=166px]{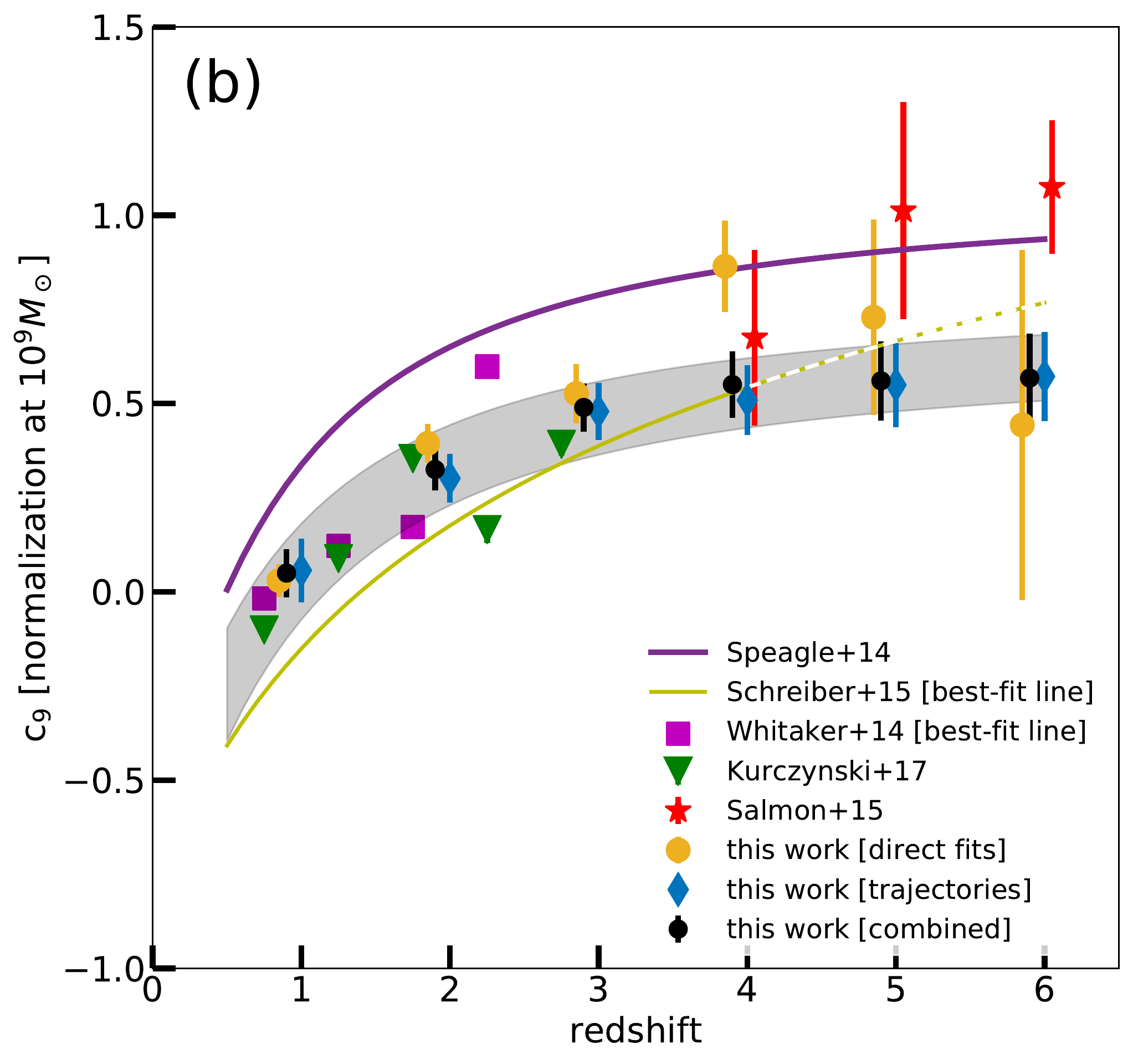}
\includegraphics[width=166px]{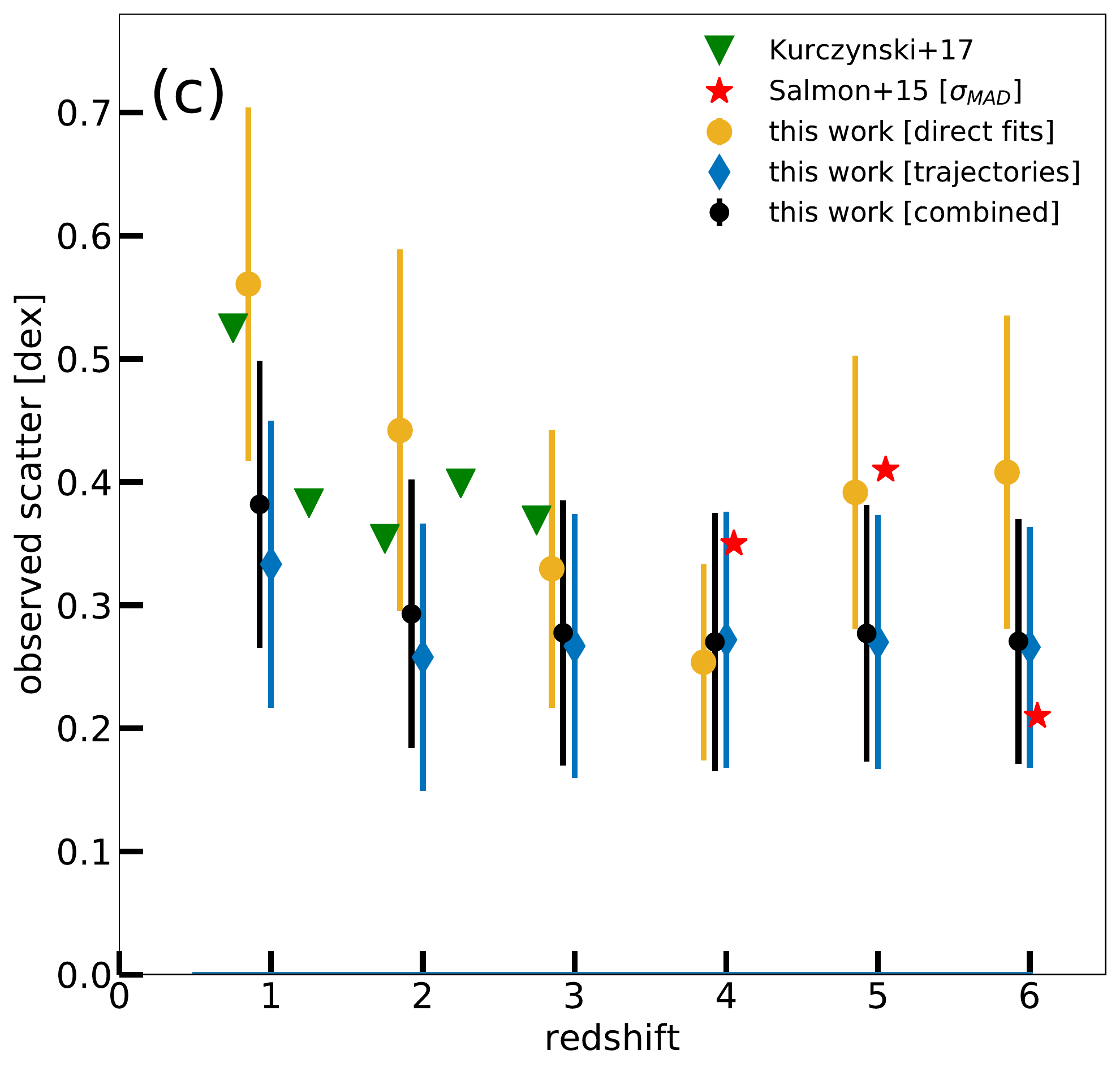}
\caption{Evolution of the slope ($m$), normalization ($c_9$, the intercept at log M$_*$ = 9) and observed scatter of the SFR-M$_*$ correlation. Black circles with errorbars show our results for the combined (direct fits + trajectories) dataset, and the grey shaded region is generated using the evolving relation defined in Eqn.\ref{eqn:2}. 
We find that the slope from the direct fits (orange circles with error bars) and trajectories (blue diamonds with error bars) are roughly consistent within uncertainties with each other at all times, with some disagreement in normalization at $2<z<4$. Our estimates for slope agree well with \citet{speagle2014highly} at low redshifts and \citet{salmon2015relation} at high redshifts, while being consistently sub-linear, in comparison to \citet{whitaker2014constraining, schreiber2015herschel}. 
\replaced{as well as the evolving \citet{speagle2014highly} relation, while the normalization follows the same trend but is systematically lower.}{Our normalization is closer to \citet{schreiber2015herschel} than \citet{speagle2014highly} in value, but has a rate of evolution more consistent with the latter. } Our measurement of observed scatter agrees well with the observed scatter reported in \citet{kurczynski2016evolution, salmon2015relation}. Measurements from  \citet{whitaker2014constraining, schreiber2015herschel} are shown for comparison, using a local slope at $10^9M_\odot$ for non-linear reported correlations. 
}
\label{fig:slope_norm_evolution}
\end{center}
\end{figure*}

\begin{table*}[ht!]
\caption{Comparing distributions from Direct fits vs Reconstructed Trajectories\footnote{For both methods (galaxies observed at a particular epoch vs those observed at lower redshifts and propagated along their trajectories), we compute the distribution of distances of individual galaxies from the combined best-fit SFR-M$_*$ correlation at z=1,2,3,4,5,6. \newline
These distributions are compared using a KS test, to test the hypothesis that they are consistent with being drawn from the same distribution \replaced{(the hypothesis fails if the p-value $> \alpha \approx 0.01$)}{We reject this hypothesis if the p-value $> \alpha (= 0.0083)$. To arrive at this value we apply a Bonferroni correction \citep{goeman2014multiple} to control for false positives since we are performing a family of tests to evaluate a single hypothesis}. \newline
Since we do not want to include starburst and quiescent galaxies, we exclude galaxies that lie at a distance $> 0.4$ dex from the correlation. \added{In Table \ref{table:KStest_diffsets} we also give p-values for the case where we exclude galaxies that lie farther than $1 \times $ the observed scatter at each redshift. The last two columns show the number of galaxies from each dataset used in comparing the two distributions.} The test shows that the direct fits and trajectories are consistent with being drawn from the same distribution at all redshifts as evinced by small values of the KS statistic, which measures the maximum distance between the CDF of the two distributions. The larger KS distance in the \replaced{last two}{$z\sim 5$} redshift bin is due to the small number of points from direct fits.\\}}
\label{table:dist_compare}
\begin{center}
\begin{tabular}{ c|c c c c c}
\hline \hline
redshift &	p-value  &	KS-statistic &	cutoff [dex] &	 \#traj &	\#direct \\
\hline
1.0	($0.9<z<1.1$)		&	0.85	&	0.03	&	0.4	&	947	&	1667 \\
	2.0	($1.9<z<2.1$)	 	& 	0.10 	&	0.06 	&	0.4	&	2950	&	1003 \\
	3.0 	($2.9<z<3.1$)	 	& 	0.02 	&	0.10 	&	0.4	&	3394	&	419 \\
	4.0 	($3.9<z<4.1$)	 	& 	0.67 	&	0.08 	&	0.4	&	3019	&	162 \\
	5.0 ($4.5<z<5.0$) 		 	& 	0.20 	&	0.13 	&	0.4	&	2396	&	183  \\
	6.0 ($5.5<z<6.0$)		 	& 	0.98 	&	0.09 	&	0.4	&	2101	&	81  \\
\hline
\end{tabular}
\end{center}
\end{table*}

To study the redshift evolution of the slope and normalization of the SFR-M$_*$ correlation, we analyze the direct fits and trajectories separately and present the results in Table \ref{table:slope_norm_evol_table} and graphically in Figure \ref{fig:slope_norm_evolution}. 
\deleted{To compare quantities on the same footing, direct fits are analyzed at $z= [1,2,3,4,5,6]$ in bins with $\Delta z = 0.2$, and the slope and normalization are determined using standard $\chi^2$ minimization.}
The slope from both approaches are consistent within uncertainties, and roughly match the published meta-analysis of \citet{speagle2014highly} in figure \ref{fig:slope_norm_evolution}, shown as a solid purple line. This result is reassuring considering the \citet{speagle2014highly} relation was calibrated at stellar masses above $10^9 M_\odot$. \added{In comparison to the \citet{speagle2014highly} relation, however, we find the slope to be consistent with little to no evolution with redshift.} The normalization for the trajectories shows the same trend in redshift as the direct fits and the \citet{speagle2014highly} relation, albeit being systematically lower at high redshifts. \added{For better comparison with literature, we also considered a set of estimates for the slope and normalization where we use an additional UVJ selection criterion to select star forming galaxies for the direct fits dataset. Details of the UVJ selection can be found in Appendix \ref{app:UVJ_preselection}, where we find that the results don't vary much due the robust fitting algorithm we use. We do not consider such a criterion for trajectories since galaxies that are quiescent at the epoch of observation can be traced back to epochs when they were star forming and thus contribute to the SFR-M$_*$ correlation at higher redshifts.} Observed scatter is computed using the same procedure as \citet{kurczynski2016evolution}, finding the standard deviation of the distribution of \replaced{distances}{$\Delta$SFR} from the best-fit correlation excluding points beyond 1 dex. \added{We find that the scatter is close to $0.3$ dex at all epochs with the possibility of being higher at low redshifts as seen for the direct fits.}  \added{The observed scatter contains contributions from noise that needs to be deconvolved to estimate the intrinsic scatter \citep{kurczynski2016evolution} and while our scatter for direct fits and trajectories are consistent within uncertainties, it is possible that we underestimate the intrinsic scatter measured with trajectories since we do not consider short-timescale excursions from the smooth best-fit SFH for individual galaxies \citep{matthee2018origin}.} The published \citet{kurczynski2016evolution} and \citet{salmon2015relation} values for slope, normalization and scatter are also shown as blue triangles at $0.5<z<2.5$ and purple stars at $3.5<z<6.5$. \citet{whitaker2014constraining} and \citet{schreiber2015herschel} report a nonlinear SFR-M$_*$ correlation due to a turnover at high stellar mass ($M_* > 10^{10.5}M_\odot$). Since most of our galaxies fall below this mass range, we fit their reported correlation near $10^9 M_\odot$ with a line to find the effective low mass slope and normalization shown in Figure \ref{fig:slope_norm_evolution}. 

Fitting the slope and normalization of the best-fit SFR-M$_*$ correlation as a function of cosmic time, we find that the relation is well described by a linear fit, given by

\begin{align}
\label{eqn:2}
\log SFR &= (0.80 \pm 0.029  - 0.017 \pm 0.010   \times  t_{univ}) \log M_* \nonumber \\    &- ( 6.487  \pm 0.282 - 0.039  \pm 0.008  \times  t_{univ})
\end{align}

where $t_{univ}$ is the age of the universe in Gyr at a given redshift. This relation is shown as the grey shaded region in Figure.\ref{fig:slope_norm_evolution}. 

\added{In estimating the slope and normalization of the SFR-M$_*$ correlation, we use all galaxies that satisfy the selection criterion described in method (c) in Appendix \ref{app:validation_contd}. In Table \ref{table:lowest_masses_probed}, we estimate the minimum well-sampled mass at each redshift, below which the statistics may be insufficient to confirm that the values for slope and normalization of the SFR-M$_*$ correlation still apply. This is based on the distribution in Stellar Mass and SFR for direct fits and trajectories, conservatively quantified as the 10th percentile of the respective distributions.  We find that we can probe the SFR-M$_*$ correlation to $\sim 1$ dex lower than possible with just direct fits. This is possible since we are no longer limited by selection effects such as the F160w detection threshold, which does not allow us to detect the faint, low mass galaxies at high redshifts that we would see at lower redshifts. Using trajectories is thus a useful tool to go deeper in SFR and M$_*$ at high redshifts.}

\added{Considering the trajectories allows us to  effectively increase the survey volume obtained by propagating galaxies observed at later epochs backwards in time. We estimate this effective increase in the volume of the survey by comparing the ratio of the number of galaxies from just direct fits vs direct fits + trajectories in a comparable mass range. The comparable mass range is obtained by requiring that the median of M$_*$ for the direct fits and trajectories in this mass range be separated by $< 0.1 dex$. In theory, this can be applied to any survey that probes a wide variety of galaxy types to allow us to further extend our SED fitting results using SFR-M$_*$ trajectories. Although it is beyond the scope of this work, it is important to include corrections to the effective volume on an individual galaxy basis (based on the amount of time they have been extrapolated backwards along their trajectory) while considering problems such as calculating luminosity functions or number densities using the combined trajectories + direct fits datasets.}

\begin{table*}[ht!]

\begin{center}
\begin{tabular}{ c|c c | c c | c}
\hline \hline
redshift & \vtop{\hbox{\strut log M$_*$ (direct)}\hbox{\strut (10th percentile)}}	 & \vtop{\hbox{\strut log M$_*$ (traj.)}\hbox{\strut (10th percentile)}} & \vtop{\hbox{\strut log SFR (direct)}\hbox{\strut (10th percentile)}}& \vtop{\hbox{\strut log SFR (traj.)}\hbox{\strut (10th percentile)}} & \vtop{\hbox{\strut Eff. Volume}\hbox{\strut (direct + traj.)}} \\
\hline
    1.0 &	7.99	&	7.62	& -1.19	&	-1.82	&	1.29x \\
	2.0	& 	7.91 	&	7.74 	& -0.85	&	-1.16	&	3.25x \\
	3.0 & 	8.59 	&	7.57 	& -0.50	&	-1.05	&	5.49x \\
	4.0 & 	8.59 	&	7.59 	& -0.40	&	-1.03	&	11.74x \\
	5.0 & 	8.67 	&	7.44 	& -1.13	&	-1.15	&	8.61x  \\
	6.0 &   8.34 	&	7.38 	& -0.13	&	-1.2	&	13.74x  \\
\hline
\end{tabular}
\end{center}
\caption{10th percentile of Stellar Mass and SFR probed using direct fits and trajectories at different redshifts. We see that using trajectories allows us to probe the SFR-M$_*$ correlation to nearly 1 dex deeper at high redshifts as compared to simply using direct fits. The last column estimates the the increase in the effective volume of the survey, using the increased number of galaxies at a particular redshift obtained by adding trajectories in a mass range where the direct fits and trajectories are comparable.}
\label{table:lowest_masses_probed}
\end{table*}

While simulations don't yet make predictions for the slope and normalization of the correlation at the lowest masses, our results help put strong constraints on the models. When comparing M$_*$, SFR distributions obtained through trajectories with simulations, it is important to k is there a physical explanation for the discrepancy between the trajectory scatter and the direct fits scatterefore, to compare SFR-M$_*$ correlations between our results and the simulations on the same footing, it is important that the correlation be compiled at any redshift using galaxies summed over all progenitors at the redshift of observation - for example, to compare accurately to reconstructed trajectories at $z\sim 6$, a simulation should be allowed to run to at least $3~Gyr$ in the future to about $z\sim 4$, then traced back to $z\sim 6$. This imposes resolution requirements at both the lower redshift, for the discovery of galaxies and at the redshift of interest, to be able to distinguish progenitors that contribute to the trajectories. The similarity of the two distributions when compared using the KS test indicate that this difference is not a major one for our observed sample of CANDELS/GOODS-S galaxies.

\begin{table*}[ht!]
\begin{center}
\begin{tabular}{ c|c c c | c c c | c c c | c}
\hline \hline
z & $m_{direct}$ & $m_{traj}$ & $m_{total}$ & $c_{9, direct}$ & $c_{9, traj}$ & $c_{9,total}$ & $\sigma_{obs, direct}$ & $\sigma_{obs, traj}$ & $\sigma_{obs, total}$ & Conf. level \\
\hline
1 & 0.65$\pm$0.06 & 0.69$\pm$0.09 & 0.69$\pm$0.08 & 0.03$\pm$0.04 & 0.06$\pm$0.08 & 0.05$\pm$0.06 & 0.56$\pm$0.14 & 0.33$\pm$0.12 & 0.38$\pm$0.12 & 91.7\% \\
2 & 0.79$\pm$0.07 & 0.75$\pm$0.07 & 0.76$\pm$0.07 & 0.39$\pm$0.05 & 0.30$\pm$0.06 & 0.32$\pm$0.06 & 0.44$\pm$0.15 & 0.26$\pm$0.11 & 0.29$\pm$0.11 & 91.2\% \\
3 & 0.79$\pm$0.10 & 0.80$\pm$0.08 & 0.80$\pm$0.07 & 0.53$\pm$0.08 & 0.48$\pm$0.08 & 0.49$\pm$0.06 & 0.33$\pm$0.11 & 0.27$\pm$0.11 & 0.28$\pm$0.11 & 98.1\% \\
4 & 0.77$\pm$0.18 & 0.76$\pm$0.09 & 0.77$\pm$0.09 & 0.86$\pm$0.12 & 0.51$\pm$0.09 & 0.55$\pm$0.09 & 0.25$\pm$0.08 & 0.27$\pm$0.10 & 0.27$\pm$0.10 & 97.9\% \\
5 & 0.70$\pm$0.26 & 0.74$\pm$0.11 & 0.75$\pm$0.10 & 0.73$\pm$0.26 & 0.55$\pm$0.11 & 0.56$\pm$0.10 & 0.39$\pm$0.11 & 0.27$\pm$0.10 & 0.28$\pm$0.10 & 94.4\% \\
6 & 0.82$\pm$0.47 & 0.78$\pm$0.11 & 0.78$\pm$0.11 & 0.44$\pm$0.47 & 0.57$\pm$0.12 & 0.57$\pm$0.12 & 0.41$\pm$0.13 & 0.27$\pm$0.10 & 0.27$\pm$0.10 & 97.6\% \\
\hline
\end{tabular}
\end{center}
\caption{Results: The slope, normalization at $10^9 M_\odot$ and observed scatter of the best-fit to the SFR-M$_*$ correlation at different redshifts as shown in Figure \ref{fig:slope_norm_evolution} for direct fits to galaxies observed at each epoch (\textbf{direct}; orange points in Figure \ref{fig:slope_norm_evolution}), galaxies observed at later epochs propagated backwards along their SFR-M$_*$ trajectories (\textbf{traj}; blue points in Figure \ref{fig:slope_norm_evolution}) and the combined sample (\textbf{total}; black points in Figure \ref{fig:slope_norm_evolution}). Including the low mass data in our fits, we observe a milder evolution of the slope and normalization with time in comparison to \citet{speagle2014highly}, finding that the evolving correlation is best described by Eqn. \ref{eqn:2}: $\log SFR = (0.80 \pm 0.029  - 0.017 \pm 0.010   \times  t_{univ}) \log M_* - ( 6.487  \pm 0.282 - 0.039  \pm 0.008  \times  t_{univ})$, where $t_{univ}$ is the age of the universe at a given redshift. The last column details the confidence levels ($ 1- $p-value) from the F-test to check the hypothesis that a linear fit to the SFR-M$_*$ is favoured over a quadratic fit.}
\label{table:slope_norm_evol_table}
\end{table*}

\section{Validation:} \label{sec:validation}

SED fitting allows us to estimate M$_{*}$ and SFRs at the epoch of observation. In addition to this, we  estimate the Stellar Masses and SFRs at previous epochs during which the galaxy was forming stars, by propagating galaxies backwards in time along their reconstructed SFR-M$_*$ trajectories. \added{In Appendix \ref{app:validation_contd} we verify the robustness of our trajectories, and restrict our analysis to the subsample of galaxies whose trajectories have low uncertainties in SFR and M$_*$ at a given redshift of interest.}  A reassuring check of the robustness of our method come from the similarity between the distributions around the SFR-M$_*$ correlation from direct fits and trajectories in Table \ref{table:dist_compare}.

\begin{table*}[ht!]
\begin{center}
\begin{tabular}{ c|c c c c c c c c}
\hline \hline
Validation test &	$m_{orig}$ & $m_{fit}$  & $c_{9,orig}$ &	$c_{9,fit}$ & $\sigma_{true}$ & $\sigma_{fit}$ &	 $\alpha_{orig} $	&	$\alpha_{fit}$ \\
\hline
Unchanged & 0.886 & 0.896$\pm$ 0.067 & -0.212 & -0.180$\pm$ 0.041 & 0.238 & 0.231 & 1 & 1 \\
Increased slope (z=1) & 1.208 & 1.189$\pm$ 0.038 & -0.135 & -0.091$\pm$ 0.040 & 0.198 & 0.213 & 1 & 1 \\
Decreased slope (z=1) & 0.478 & 0.461$\pm$ 0.072 & -0.302 & -0.288$\pm$ 0.038 & 0.231 & 0.217 & 1 & 1 \\
Changed shape (z=1)$^\dagger$ & -0.075 & -0.086$\pm$ 0.093 & -0.215 & -0.228$\pm$ 0.142 & 0.607 & 0.629 & 2 & 2 \\
Increased slope (z=2) & 1.049 & 1.028$\pm$ 0.015 & -0.158 & 0.082$\pm$ 0.054 & 0.249 & 0.403 & 1 & 1 \\
Decreased slope (z=2) & 0.663 & 0.532$\pm$ 0.055 & 0.165 & 0.268$\pm$ 0.027 & 0.059 & 0.181 & 1 & 1 \\
Changed shape (z=2)$^\dagger$ & 0.459 & 0.495$\pm$ 0.049 & -0.421 & -0.139$\pm$ 0.067 & 0.640 & 0.389 & 2 & 2 \\
\hline
\end{tabular}
\end{center}
\caption{ Validation: estimating the sensitivity of the fits to slope, normalization and shape of the SFR-M$_*$ correlation. $m$ is the slope of the linear correlation, $c$ is the normalization, and $\alpha$ denotes the degree of the polynomial that the correlation is best fit with. The $m_{orig}$ and $c_{orig}$ are the linear coefficients obtained from the best-fit to the SFR-M$_*$ correlation  generated directly using SED fitting with the modified SFHs for each test case. \\
$^\dagger$Although the slope and normalization are reported, the correlation in this case is better fit with a quadratic, and is not well described by the linear coefficients. For the $z=1$ case, the true correlation is described by $\log$ SFR $= -0.364 (log M_* -9)^2 + 0.515 (log M_* -9) +0.053 $. For the $z=2$ case, the coefficients are log SFR $= -0.199 (log M_*-9)^2 + 0.342 (log M_*-9) +0.271 $.  }
\label{table:slope_changes}
\end{table*}

To further ensure that we do not get an artificially linear correlation due to our fitting method, we use a sample of mock SFHs from the MUFASA hydrodynamic simulations \citep{dave2016mufasa} to run a series of validation tests by changing the slope, normalization and shape of the simulated SFR-M$_*$ correlation to see if our fits can recover these changes. The results of are reported in Table \ref{table:slope_changes}. The first six columns report the results of a linear fit, to $\log SFR = m \log [M_*/10^9M_\odot] + c_9$, reporting the slope ($m$), normalization ($c_9$) and the observed scatter ($\sigma$). The last two columns compare the goodness of fit of a first order (linear) and second order (quadratic) polynomial fit using an F-test, to determine if the improvement upon fitting with a second-order curve is statistically significant. This test allows us to test the hypothesis that the correlation is linear. For all tests, the simulated galaxies are fit at $z=1$.

The first row of Table \ref{table:slope_changes} evaluates that the SFR-M$_*$ correlation is robustly recovered for a randomly selected sample of galaxies at the epoch of observation ($z=1$, direct fits) with no modifications to their SFHs. This represents the control case for our validation tests. The next two rows change the slope of the correlation to see if our method can recover the artificially high or low slope. This is done using a gaussian envelope to modify the SFH in a way that the overall galaxy mass remains constant. We find that the recovered slope and normalization match the truth within uncertainties, as seen in columns 2-5 of Table.\ref{table:slope_changes}. In the fourth row  we check that the recovered correlation is sensitive to the linearity of the underlying correlation by curving the correlation such that it is better fit by a quadratic curve. This is achieved by adding a single gaussian component to a randomly drawn MUFASA SFH while varying  lookback time at which the SFH peaks. This results in a curved SFR-M$_*$ correlation. Although there is always an improvement to the fit with an additional degree of freedom, we use an F-test with a threshold p-value of 0.1 to see if the quadratic fit provides a statistically significant improvement in describing the variance of the data and find that the order of our recovered correlation matches the input, shown in columns 8-9 of Table.\ref{table:slope_changes}.

To check the robustness of the SFR-M$_*$ correlation recovered from galaxies propagated backwards in time along their trajectories, we also perform the same tests using a randomly selected sample of galaxies that are fit at $z=1$ and analyzed at $z=2$. We choose $z=2$ to test the trajectories, since we find that most galaxies contributing to the SFR-M$_*$ correlation through trajectories at different redshifts form the bulk of their stellar mass at $\leq 3Gyr$ from the epoch of interest. We repeat the same tests for the trajectories as we did for the direct fits, while attempting to keep the z=1 correlation the same. Since achieving such changes through modifications to the MUFASA SFHs was difficult, we use a Monte Carlo method generating random gaussian contributions to the SFH until the resulting SFH satisfies a `normal' MS slope at $z=1$ and a modified slope at $z=2$ significantly higher or lower than $m \sim 0.89$. We then generate SEDs corresponding to these SFHs and fit them, finding that the fitting method is reasonably sensitive to changes in the slope of the correlation at higher redshifts. A small number of quiescent galaxies at $z=1$ also contribute to the correlation at higher redshifts, which helps increase the sensitivity of our approach. Additionally, we repeated the test where the shape of the correlation is changed to being better described by a quadratic rather than a linear curve, finding that this too is robustly recovered by the fits.

\section{Discussion} \label{sec:systematics}

\added{The tight correlation between the SFRs and Stellar Mass of star forming galaxies has been extensively studied, with simulations matching observations at z$\sim$0 and high redshifts \citep{sparre2015star, salmon2015relation}, but with some tension at intermediate redshifts around $z \sim 2$ \citep{sparre2015star}. 
The Dense Basis method allows us to extend the dynamic range across which we fit the SFR-M$_*$ correlation to estimate the slope and normalization, helping provide more robust estimates of these quantities. 
We find sub-linear slopes at all redshifts consistent with mild evolution, similar to \citet{speagle2014highly, salmon2015relation} and \citet{kurczynski2016evolution} at $z>1$. This is in contrast (about 2.5 $\sigma$) to \citet{schreiber2015herschel} and \citet{whitaker2012star}, who find a slope closer to 1. \citet{speagle2014highly} attributes the slope to a steady, environment driven mode of star formation, where a slope slightly below unity occurs due to feedback, leading to the growth of hot halos around higher mass galaxies and slows down gas accretion \citep{finlator2006physical, dave2008galaxy}. 
\citet{salmon2015relation} argues that gas accretion onto dark-matter halos at high-z is smooth over large timescales \citep{cattaneo2011galaxies, finlator2011smoothly} assuming a power-law form of the SFH, which controls the scaling of both the SFR and stellar mass \citep{stark2009evolutionary, gonzalez2011evolution, papovich2011rising}. Results from our more versatile SFHs appear to extend this interpretation across a wider range of redshifts. 
A sub-linear slope to the SFR-M$_*$ correlation is also relevant in the context of \citet{abramson2016return}, which considers how the growth of a bulge adds M$_*$ but not SFR. It would be an interesting analysis to further study how the scatter around the SFR-M$_*$ correlation correlates with explicit SFH parameters like t10 (the lookback time when the galaxy forms the first $10\%$ of its stellar mass) and morphological quantities like the bulge/disk ratio \citep{abramson2014mass}. 
The evolution of the normalization, which could be related to changing cosmological gas accretion rates with redshift \citep{dutton2010origin} generally agrees with the literature, while being $\sim 0.2-0.4$ dex lower than the \citet{speagle2014highly} meta-analysis at all redshifts and $\sim 0.5$ dex lower than \citet{salmon2015relation} at $z\sim 6$, albeit with larger uncertainties.}

SFR-M$_*$ trajectories obtained through SED fitting provide a valuable tool to extract information about where galaxies lie on the Stellar Mass - SFR plane at multiple epochs, allowing us to probe the low-mass portion of the SFR-M$_*$ correlation as we go to higher redshifts, with greater numbers than previously available. However, in interpreting the results we obtain, it is important to keep in mind the limitations of the observational data, as well as the current implementation of the Dense Basis method. Since our method estimates the smooth overall trend of star formation in a galaxy's past, it does not recover stochastic `short timescale' star formation events that contribute to the intrinsic scatter of the SFR-M$_*$ correlation. However, we find that only  $\sim$5\% of a sample of galaxies from SAMs \citep{somerville2015star} and $\sim$7\% of galaxies from MUFASA \citep{dave2016mufasa} show a short burst where SFR$_{10Myr}$ / SFR$_{life} >$ 10, where SFR$_{10Myr}$ is the SFR averaged over the last 10 Myr lookback time, and SFR$_{life}$ is the SFR of a galaxy averaged over its lifetime. This is not significant enough to bias estimations of the slope/normalization, or to alter the results from our validation tests.

In hierarchical cosmology, the process of galaxy evolution includes major and minor mergers along with gradual mass growth due to infall.
When SED fitting yields information about an observed galaxy's star formation history, even a summary statistic such as the age of its stellar population, this information is about the sum of all stars in that galaxy’s progenitors.
Hence the trajectories that result from our SFH reconstruction represent trajectories of the summed stellar masses and star formation rates of each observed galaxy's progenitors, rather than those of its most massive progenitor at each epoch.
However, observable samples of galaxies at e.g., $z=4$, will contain the most massive progenitors of observable galaxies at e.g, $z=2$, along with a few additional progenitors that are massive enough to be detected.  
Looking backwards in time, a minor merger with 10:1 or 3:1 mass ratio causes only a 0.04  or 0.12 dex offset, respectively, in mass between the sum-of-progenitors and the most massive progenitor, with the maximum offset of 0.3 dex coming from a 1:1 major merger.  
Such major mergers are predicted to be rare \citep{kaviraj2015galaxy, rodriguez2015merger, ventou2017muse}, even at high redshift, for galaxies massive enough to be detected at $z\sim1$.  
Nonetheless, the KS test described in \S \ref{sec:results} found the distribution of trajectory values about the inferred SFR-M$_*$ correlation to match that of the observed values from direct fits well enough that the hypothesis of these being drawn from a single underlying population is not ruled out at higher than 99\% confidence for any of the samples.

Recent studies \citep{hsieh2017sdss} indicate that a strong correlation exists between the Star Formation Rate Density ($\Sigma_{SFR}$) and the Surface Mass Density ($\Sigma_{M_*}$) in star forming galaxies at kpc scales. This indicates that the SFR-M$_*$ correlation may extend to much lower scales than currently measured, with the method described in this work providing a unique bridge to intermediate scales.

\section{Conclusions} \label{sec:finish}

SFH reconstruction through SED fitting yields the trajectories of galaxies that evolve through SFR-M$_*$ space and is thus uniquely suited to probe the low-mass end of the SFR-M$_*$ correlation as we go to higher redshifts.
In this paper, we used the Dense Basis method \citep{iyer2017reconstruction} to fit a sample of $\sim$17,800 galaxies in the CANDELS GOODS-S field at redshifts $0.5<z<6.0$. 
We used the reconstructed SFHs to obtain the stellar masses and star formation rates of galaxies at $z = 1,2,3,4,5,6$. Using the combined dataset from galaxies observed at the epochs of interest (direct fits) and galaxies observed at lower epochs propagated backwards in time along their SFR-M$_*$ trajectories, we find that the SFR-M$_*$ correlation is linear to $\sim 10^6 M_\odot$ at high redshifts.

This allows us to study the nature and evolution of the SFR-M$_*$ correlation in greater detail than allowed by previous approaches like Main Sequence Integration \citep{leitnerMSI} which assumes that star forming galaxies stay on the correlation throughout their lifetimes. We find that the overall trend of the evolution of the slope of the SFR-M$_*$ correlation with redshift is roughly consistent with the evolving \citet{speagle2014highly} relation, while the normalization seems systematically lower by a factor of $\sim 0.2$ dex. 

Thus new approach provides a probe of the correlation at much lower masses than previously possible
\citep{salmon2015relation, kurczynski2016evolution}
since stellar masses decrease as we propagate galaxies backwards along their SFR-M$_*$ trajectories. This is more important in view of the selection effects in the direction of increasing stellar mass as we go to higher redshift in galaxy surveys like CANDELS.
It also allows for a closer comparison between observations and simulations, by providing constraints for simulations through the comparison of the predicted SFR-M$_*$ distributions to the reconstructed ones down to much lower masses.

\section*{Acknowledgements}

The authors would like to thank \added{the anonymous referee, in addition to } Louis Abramson, Matt Ashby, Nimish Hathi, Kameswara Bharadwaj Mantha, Mara Salvato, Chris Lovell and Adriano Fontana for their insightful comments and suggestions.
The authors acknowledge Steve Finkelstein, Adriano Fontana, Janine Pforr, Mara Salvato, Tommy Wiklind, and Stijn Wuyts for generating the photo-z PDFs for the compilation of $z_{\rm best}$ in the v2 CANDELS photo-z catalog used in this work. KI \& EG gratefully acknowledge support from Rutgers University. 
This work used resources from the Rutgers Discovery Informatics Institute, supported by Rutgers and the State of New Jersey. The Flatiron Institute is supported by the Simons Foundation.
Support for Program number HST-AR-14564.001-A was provided by NASA through a grant from the Space Telescope Science Institute, which is operated by the Association of Universities for Research in Astronomy, Incorporated, under NASA contract NAS5-26555.

\bibliography{db_refs.bib} 

\begin{thebibliography}{}
\expandafter\ifx\csname natexlab\endcsname\relax\def\natexlab#1{#1}\fi

\bibitem[{Abramson {et~al.}(2016)Abramson, Gladders, Dressler, Oemler~Jr,
  Poggianti, \& Vulcani}]{abramson2016return}
Abramson, L.~E., Gladders, M.~D., Dressler, A., {et~al.} 2016, The
  Astrophysical Journal, 832, 7

\bibitem[{Abramson {et~al.}(2014)Abramson, Kelson, Dressler, Poggianti,
  Gladders, Oemler~Jr, \& Vulcani}]{abramson2014mass}
Abramson, L.~E., Kelson, D.~D., Dressler, A., {et~al.} 2014, The Astrophysical
  Journal Letters, 785, L36

\bibitem[{Acquaviva {et~al.}(2011)Acquaviva, Gawiser, \&
  Guaita}]{acquaviva2011sed}
Acquaviva, V., Gawiser, E., \& Guaita, L. 2011, Proceedings of the
  International Astronomical Union, 7, 42

\bibitem[{Ashby {et~al.}(2015)Ashby, Willner, Fazio, Dunlop, Egami, Faber,
  Ferguson, Grogin, Hora, Huang, {et~al.}}]{ashby2015s}
Ashby, M., Willner, S., Fazio, G., {et~al.} 2015, The Astrophysical Journal
  Supplement Series, 218, 33

\bibitem[{Calzetti(2001)}]{calzetti2001dust}
Calzetti, D. 2001, Publications of the Astronomical Society of the Pacific,
  113, 1449

\bibitem[{Cattaneo {et~al.}(2011)Cattaneo, Mamon, Warnick, \&
  Knebe}]{cattaneo2011galaxies}
Cattaneo, A., Mamon, G.~A., Warnick, K., \& Knebe, A. 2011, Astronomy \&
  Astrophysics, 533, A5

\bibitem[{Chabrier(2003)}]{chabrier2003galactic}
Chabrier, G. 2003, Publications of the Astronomical Society of the Pacific,
  115, 763

\bibitem[{Ciesla {et~al.}(2017)Ciesla, Elbaz, \& Fensch}]{ciesla2017sfr}
Ciesla, L., Elbaz, D., \& Fensch, J. 2017, arXiv preprint arXiv:1706.08531

\bibitem[{Cleveland(1979)}]{loess}
Cleveland, W.~S. 1979, Journal of the American statistical association, 74, 829

\bibitem[{Conroy \& Gunn(2010)}]{conroy2010propagation}
Conroy, C., \& Gunn, J.~E. 2010, The Astrophysical Journal, 712, 833

\bibitem[{Conroy {et~al.}(2009)Conroy, Gunn, \& White}]{conroy2009propagation}
Conroy, C., Gunn, J.~E., \& White, M. 2009, The Astrophysical Journal, 699, 486

\bibitem[{Daddi {et~al.}(2007)Daddi, Dickinson, Morrison, Chary, Cimatti,
  Elbaz, Frayer, Renzini, Pope, Alexander, {et~al.}}]{daddi2007multiwavelength}
Daddi, E., Dickinson, M., Morrison, G., {et~al.} 2007, The Astrophysical
  Journal, 670, 156

\bibitem[{Dav{\'e}(2008)}]{dave2008galaxy}
Dav{\'e}, R. 2008, Monthly Notices of the Royal Astronomical Society, 385, 147

\bibitem[{Dav{\'e} {et~al.}(2016)Dav{\'e}, Thompson, \&
  Hopkins}]{dave2016mufasa}
Dav{\'e}, R., Thompson, R., \& Hopkins, P.~F. 2016, Monthly Notices of the
  Royal Astronomical Society, 462, 3265

\bibitem[{Dutton {et~al.}(2010)Dutton, van~den Bosch, \&
  Dekel}]{dutton2010origin}
Dutton, A.~A., van~den Bosch, F.~C., \& Dekel, A. 2010, Monthly Notices of the
  Royal Astronomical Society, 405, 1690

\bibitem[{Elbaz {et~al.}(2007)Elbaz, Daddi, Le~Borgne, Dickinson, Alexander,
  Chary, Starck, Brandt, Kitzbichler, MacDonald, {et~al.}}]{elbaz2007reversal}
Elbaz, D., Daddi, E., Le~Borgne, D., {et~al.} 2007, Astronomy \& Astrophysics,
  468, 33

\bibitem[{Finlator {et~al.}(2006)Finlator, Dav{\'e}, Papovich, \&
  Hernquist}]{finlator2006physical}
Finlator, K., Dav{\'e}, R., Papovich, C., \& Hernquist, L. 2006, The
  Astrophysical Journal, 639, 672

\bibitem[{Finlator {et~al.}(2011)Finlator, Oppenheimer, \&
  Dav{\'e}}]{finlator2011smoothly}
Finlator, K., Oppenheimer, B.~D., \& Dav{\'e}, R. 2011, Monthly Notices of the
  Royal Astronomical Society, 410, 1703

\bibitem[{Fontana {et~al.}(2014)Fontana, Dunlop, Paris, Targett, Boutsia,
  Castellano, Galametz, Grazian, McLure, Merlin, {et~al.}}]{fontana2014hawk}
Fontana, A., Dunlop, J., Paris, D., {et~al.} 2014, Astronomy \& Astrophysics,
  570, A11

\bibitem[{Forbes {et~al.}(2014)Forbes, Krumholz, Burkert, \&
  Dekel}]{forbes2014origin}
Forbes, J.~C., Krumholz, M.~R., Burkert, A., \& Dekel, A. 2014, Monthly Notices
  of the Royal Astronomical Society, 443, 168

\bibitem[{Foreman-Mackey {et~al.}(2014)Foreman-Mackey, Sick, \&
  Johnson}]{dan_foreman_mackey_2014_12157}
Foreman-Mackey, D., Sick, J., \& Johnson, B. 2014, python-fsps: Python bindings
  to FSPS (v0.1.1), doi:10.5281/zenodo.12157

\bibitem[{Galametz {et~al.}(2013)Galametz, Grazian, Fontana, Ferguson, Ashby,
  Barro, Castellano, Dahlen, Donley, Faber, {et~al.}}]{galametz2013candels}
Galametz, A., Grazian, A., Fontana, A., {et~al.} 2013, The Astrophysical
  Journal Supplement Series, 206, 10

\bibitem[{Goeman \& Solari(2014)}]{goeman2014multiple}
Goeman, J.~J., \& Solari, A. 2014, Statistics in medicine, 33, 1946

\bibitem[{Gonz{\'a}lez {et~al.}(2011)Gonz{\'a}lez, Labb{\'e}, Bouwens,
  Illingworth, Franx, \& Kriek}]{gonzalez2011evolution}
Gonz{\'a}lez, V., Labb{\'e}, I., Bouwens, R.~J., {et~al.} 2011, The
  Astrophysical Journal Letters, 735, L34

\bibitem[{Grogin {et~al.}(2011)Grogin, Kocevski, Faber, Ferguson, Koekemoer,
  Riess, Acquaviva, Alexander, Almaini, Ashby, {et~al.}}]{candels}
Grogin, N.~A., Kocevski, D.~D., Faber, S., {et~al.} 2011, The Astrophysical
  Journal Supplement Series, 197, 35

\bibitem[{Guo {et~al.}(2013)Guo, Ferguson, Giavalisco, Barro, Willner, Ashby,
  Dahlen, Donley, Faber, Fontana, {et~al.}}]{guo2013candels}
Guo, Y., Ferguson, H.~C., Giavalisco, M., {et~al.} 2013, The Astrophysical
  Journal Supplement Series, 207, 24

\bibitem[{Holland \& Welsch(1977)}]{holland1977robust}
Holland, P.~W., \& Welsch, R.~E. 1977, Communications in Statistics-theory and
  Methods, 6, 813

\bibitem[{Hsieh {et~al.}(2017)Hsieh, Lin, Lin, Pan, Hsu, S{\'a}nchez,
  Cano-D{\'\i}az, Zhang, Yan, Barrera-Ballesteros, {et~al.}}]{hsieh2017sdss}
Hsieh, B., Lin, L., Lin, J., {et~al.} 2017, The Astrophysical Journal Letters,
  851, L24

\bibitem[{Hsu {et~al.}(2014)Hsu, Salvato, Nandra, Brusa, Bender, Buchner,
  Donley, Kocevski, Guo, Hathi, {et~al.}}]{hsu2014candels}
Hsu, L.-T., Salvato, M., Nandra, K., {et~al.} 2014, The Astrophysical Journal,
  796, 60

\bibitem[{Iyer \& Gawiser(2017)}]{iyer2017reconstruction}
Iyer, K., \& Gawiser, E. 2017, The Astrophysical Journal, 838, 127

\bibitem[{Johnston {et~al.}(2015)Johnston, Vaccari, Jarvis, Smith, Giovannoli,
  H{\"a}u{\ss}ler, \& Prescott}]{johnston2015evolving}
Johnston, R., Vaccari, M., Jarvis, M., {et~al.} 2015, Monthly Notices of the
  Royal Astronomical Society, 453, 2540

\bibitem[{Kaviraj {et~al.}(2015)Kaviraj, Devriendt, Dubois, Slyz, Welker,
  Pichon, Peirani, \& Le~Borgne}]{kaviraj2015galaxy}
Kaviraj, S., Devriendt, J., Dubois, Y., {et~al.} 2015, Monthly Notices of the
  Royal Astronomical Society, 452, 2845

\bibitem[{Koekemoer {et~al.}(2011)Koekemoer, Faber, Ferguson, Grogin, Kocevski,
  Koo, Lai, Lotz, Lucas, McGrath, {et~al.}}]{koekemoer2011candels}
Koekemoer, A.~M., Faber, S., Ferguson, H.~C., {et~al.} 2011, The Astrophysical
  Journal Supplement Series, 197, 36

\bibitem[{Kurczynski {et~al.}(2016)Kurczynski, Gawiser, Acquaviva, Bell, Dekel,
  de~Mello, Ferguson, Gardner, Grogin, Guo, {et~al.}}]{kurczynski2016evolution}
Kurczynski, P., Gawiser, E., Acquaviva, V., {et~al.} 2016, The Astrophysical
  Journal Letters, 820, L1

\bibitem[{Laidler {et~al.}(2007)Laidler, Papovich, Grogin, Idzi, Dickinson,
  Ferguson, Hilbert, Clubb, \& Ravindranath}]{laidler2007tfit}
Laidler, V.~G., Papovich, C., Grogin, N.~A., {et~al.} 2007, Publications of the
  Astronomical Society of the Pacific, 119, 1325

\bibitem[{Lee {et~al.}(2017)Lee, Giavalisco, Whitaker, Williams, Ferguson,
  Acquaviva, Koekemoer, Straughn, Guo, Kartaltepe, {et~al.}}]{lee2017intrinsic}
Lee, B., Giavalisco, M., Whitaker, K., {et~al.} 2017, arXiv preprint
  arXiv:1706.02311

\bibitem[{Leitner(2012)}]{leitnerMSI}
Leitner, S.~N. 2012, The Astrophysical Journal, 745, 149

\bibitem[{Lilly {et~al.}(2013)Lilly, Carollo, Pipino, Renzini, \&
  Peng}]{lilly2013gas}
Lilly, S.~J., Carollo, C.~M., Pipino, A., Renzini, A., \& Peng, Y. 2013, The
  Astrophysical Journal, 772, 119

\bibitem[{Madau {et~al.}(1996)Madau, Ferguson, Dickinson, Giavalisco, Steidel,
  \& Fruchter}]{madau1996high}
Madau, P., Ferguson, H.~C., Dickinson, M.~E., {et~al.} 1996, Monthly Notices of
  the Royal Astronomical Society, 283, 1388

\bibitem[{Mannucci {et~al.}(2010)Mannucci, Cresci, Maiolino, Marconi, \&
  Gnerucci}]{mannucci2010fundamental}
Mannucci, F., Cresci, G., Maiolino, R., Marconi, A., \& Gnerucci, A. 2010,
  Monthly Notices of the Royal Astronomical Society, 408, 2115

\bibitem[{Matthee \& Schaye(2018)}]{matthee2018origin}
Matthee, J., \& Schaye, J. 2018, arXiv preprint arXiv:1805.05956

\bibitem[{Mitra {et~al.}(2016)Mitra, Dav{\'e}, Simha, \&
  Finlator}]{mitra2016equilibrium}
Mitra, S., Dav{\'e}, R., Simha, V., \& Finlator, K. 2016, Monthly Notices of
  the Royal Astronomical Society, 464, 2766

\bibitem[{Mu{\~n}oz \& Peeples(2015)}]{munoz2015framework}
Mu{\~n}oz, J.~A., \& Peeples, M.~S. 2015, Monthly Notices of the Royal
  Astronomical Society, 448, 1430

\bibitem[{Noeske {et~al.}(2007)Noeske, Weiner, Faber, Papovich, Koo,
  Somerville, Bundy, Conselice, Newman, Schiminovich,
  {et~al.}}]{noeske2007star}
Noeske, K., Weiner, B., Faber, S., {et~al.} 2007, The Astrophysical Journal
  Letters, 660, L43

\bibitem[{Nonino {et~al.}(2009)Nonino, Dickinson, Rosati, Grazian, Reddy,
  Cristiani, Giavalisco, Kuntschner, Vanzella, Daddi,
  {et~al.}}]{nonino2009deep}
Nonino, M., Dickinson, M., Rosati, P., {et~al.} 2009, The Astrophysical Journal
  Supplement Series, 183, 244

\bibitem[{Pacifici {et~al.}(2012)Pacifici, Kassin, Weiner, Charlot, \&
  Gardner}]{pacifici}
Pacifici, C., Kassin, S.~A., Weiner, B., Charlot, S., \& Gardner, J.~P. 2012,
  The Astrophysical Journal Letters, 762, L15

\bibitem[{Pacifici {et~al.}(2016)Pacifici, Oh, Oh, Lee, \&
  Yi}]{pacifici2016timing}
Pacifici, C., Oh, S., Oh, K., Lee, J., \& Yi, S.~K. 2016, arXiv preprint
  arXiv:1604.02460

\bibitem[{Papovich {et~al.}(2011)Papovich, Finkelstein, Ferguson, Lotz, \&
  Giavalisco}]{papovich2011rising}
Papovich, C., Finkelstein, S.~L., Ferguson, H.~C., Lotz, J.~M., \& Giavalisco,
  M. 2011, Monthly Notices of the Royal Astronomical Society, 412, 1123

\bibitem[{Rasmussen \& Williams(2006)}]{gp_book}
Rasmussen, C.~E., \& Williams, C.~K. 2006, The MIT Press, Cambridge, MA, USA,
  38, 715

\bibitem[{Retzlaff {et~al.}(2010)Retzlaff, Rosati, Dickinson, Vandame,
  Rit{\'e}, Nonino, \& Cesarsky}]{retzlaff2010great}
Retzlaff, J., Rosati, P., Dickinson, M., {et~al.} 2010, Astronomy \&
  Astrophysics, 511, A50

\bibitem[{Rodriguez-Gomez {et~al.}(2015)Rodriguez-Gomez, Genel, Vogelsberger,
  Sijacki, Pillepich, Sales, Torrey, Snyder, Nelson, Springel,
  {et~al.}}]{rodriguez2015merger}
Rodriguez-Gomez, V., Genel, S., Vogelsberger, M., {et~al.} 2015, Monthly
  Notices of the Royal Astronomical Society, 449, 49

\bibitem[{Rodr{\'\i}guez-Puebla {et~al.}(2015)Rodr{\'\i}guez-Puebla, Primack,
  Behroozi, \& Faber}]{rodriguez2015main}
Rodr{\'\i}guez-Puebla, A., Primack, J.~R., Behroozi, P., \& Faber, S. 2015,
  Monthly Notices of the Royal Astronomical Society, 455, 2592

\bibitem[{Salim {et~al.}(2007)Salim, Rich, Charlot, Brinchmann, Johnson,
  Schiminovich, Seibert, Mallery, Heckman, Forster, {et~al.}}]{salim2007uv}
Salim, S., Rich, R.~M., Charlot, S., {et~al.} 2007, The Astrophysical Journal
  Supplement Series, 173, 267

\bibitem[{Salmon {et~al.}(2015)Salmon, Papovich, Finkelstein, Tilvi, Finlator,
  Behroozi, Dahlen, Dav{\'e}, Dekel, Dickinson, {et~al.}}]{salmon2015relation}
Salmon, B., Papovich, C., Finkelstein, S.~L., {et~al.} 2015, The Astrophysical
  Journal, 799, 183

\bibitem[{Santini {et~al.}(2015)Santini, Ferguson, Fontana, Mobasher, Barro,
  Castellano, Finkelstein, Grazian, Hsu, Lee, {et~al.}}]{santini2015stellar}
Santini, P., Ferguson, H., Fontana, A., {et~al.} 2015, The Astrophysical
  Journal, 801, 97

\bibitem[{Santini {et~al.}(2017)Santini, Fontana, Castellano, Di~Criscienzo,
  Merlin, Amorin, Cullen, Daddi, Dickinson, Dunlop, {et~al.}}]{santini2017star}
Santini, P., Fontana, A., Castellano, M., {et~al.} 2017, The Astrophysical
  Journal, 847, 76

\bibitem[{Schreiber {et~al.}(2015)Schreiber, Pannella, Elbaz, B{\'e}thermin,
  Inami, Dickinson, Magnelli, Wang, Aussel, Daddi,
  {et~al.}}]{schreiber2015herschel}
Schreiber, C., Pannella, M., Elbaz, D., {et~al.} 2015, Astronomy \&
  Astrophysics, 575, A74

\bibitem[{{\v{S}}id{\'a}k(1967)}]{vsidak1967rectangular}
{\v{S}}id{\'a}k, Z. 1967, Journal of the American Statistical Association, 62,
  626

\bibitem[{Somerville {et~al.}(2008)Somerville, Hopkins, Cox, Robertson, \&
  Hernquist}]{somerville2008semi}
Somerville, R.~S., Hopkins, P.~F., Cox, T.~J., Robertson, B.~E., \& Hernquist,
  L. 2008, Monthly Notices of the Royal Astronomical Society, 391, 481

\bibitem[{Somerville {et~al.}(2015)Somerville, Popping, \&
  Trager}]{somerville2015star}
Somerville, R.~S., Popping, G., \& Trager, S.~C. 2015, Monthly Notices of the
  Royal Astronomical Society, 453, 4337

\bibitem[{Sparre {et~al.}(2015)Sparre, Hayward, Springel, Vogelsberger, Genel,
  Torrey, Nelson, Sijacki, \& Hernquist}]{sparre2015star}
Sparre, M., Hayward, C.~C., Springel, V., {et~al.} 2015, Monthly Notices of the
  Royal Astronomical Society, 447, 3548

\bibitem[{Speagle {et~al.}(2014)Speagle, Steinhardt, Capak, \&
  Silverman}]{speagle2014highly}
Speagle, J.~S., Steinhardt, C.~L., Capak, P.~L., \& Silverman, J.~D. 2014, The
  Astrophysical Journal Supplement Series, 214, 15

\bibitem[{Stark {et~al.}(2009)Stark, Ellis, Bunker, Bundy, Targett, Benson, \&
  Lacy}]{stark2009evolutionary}
Stark, D.~P., Ellis, R.~S., Bunker, A., {et~al.} 2009, The Astrophysical
  Journal, 697, 1493

\bibitem[{Steinhardt {et~al.}(2014)Steinhardt, Speagle, Capak, Silverman,
  Carollo, Dunlop, Hashimoto, Hsieh, Ilbert, Le~Fevre,
  {et~al.}}]{steinhardt2014star}
Steinhardt, C.~L., Speagle, J.~S., Capak, P., {et~al.} 2014, The Astrophysical
  journal letters, 791, L25

\bibitem[{Tacchella {et~al.}(2016)Tacchella, Dekel, Carollo, Ceverino, DeGraf,
  Lapiner, Mandelker, \& Primack~Joel}]{tacchella2016confinement}
Tacchella, S., Dekel, A., Carollo, C.~M., {et~al.} 2016, Monthly Notices of the
  Royal Astronomical Society, 457, 2790

\bibitem[{Tasca {et~al.}(2015)Tasca, Le~F{\`e}vre, Hathi, Schaerer, Ilbert,
  Zamorani, Lemaux, Cassata, Garilli, Le~Brun, {et~al.}}]{tasca2015evolving}
Tasca, L., Le~F{\`e}vre, O., Hathi, N., {et~al.} 2015, Astronomy \&
  Astrophysics, 581, A54

\bibitem[{Torrey {et~al.}(2017)Torrey, Vogelsberger, Hernquist, McKinnon,
  Marinacci, Simcoe, Springel, Pillepich, Naiman, Pakmor,
  {et~al.}}]{torrey2017similar}
Torrey, P., Vogelsberger, M., Hernquist, L., {et~al.} 2017, arXiv preprint
  arXiv:1711.11039

\bibitem[{Ventou {et~al.}(2017)Ventou, Contini, Bouch{\'e}, Epinat, Brinchmann,
  Bacon, Inami, Lam, Drake, Garel, {et~al.}}]{ventou2017muse}
Ventou, E., Contini, T., Bouch{\'e}, N., {et~al.} 2017, Astronomy \&
  Astrophysics, 608, A9

\bibitem[{Whitaker {et~al.}(2012)Whitaker, Van~Dokkum, Brammer, \&
  Franx}]{whitaker2012star}
Whitaker, K.~E., Van~Dokkum, P.~G., Brammer, G., \& Franx, M. 2012, The
  Astrophysical Journal Letters, 754, L29

\bibitem[{Whitaker {et~al.}(2014)Whitaker, Franx, Leja, Van~Dokkum, Henry,
  Skelton, Fumagalli, Momcheva, Brammer, Labb{\'e},
  {et~al.}}]{whitaker2014constraining}
Whitaker, K.~E., Franx, M., Leja, J., {et~al.} 2014, The Astrophysical Journal,
  795, 104

\bibitem[{Williams {et~al.}(2009)Williams, Quadri, Franx, Van~Dokkum, \&
  Labb{\'e}}]{williams2009detection}
Williams, R.~J., Quadri, R.~F., Franx, M., Van~Dokkum, P., \& Labb{\'e}, I.
  2009, The Astrophysical Journal, 691, 1879

\end{thebibliography}



\appendix

\section{Validation of trajectory robustness}
\label{app:validation_contd}

In Sec. \ref{sec:validation}, we showed that our SED fitting technique is robustly able to recover the SFR-M$_*$ correlation corresponding to a mock dataset from the MUFASA simulation. This is true both for the case of direct fits at $z \sim 1$, and for stellar masses and star formation rates recovered from the SFR-M$_*$ trajectories at $z \sim 2$. 

Here we check for biases in the reconstruction of individual trajectories in SFR-M$_*$ space at different redshifts. If the estimated uncertainty from our SED fit is bigger than the bias, this should not affect our analysis since the corresponding uncertainties will re-weight these points when we fit for the slope and normalization. We also examine possible ways where we can reduce any possible ensemble bias using tracers that are sensitive to factors like the rest-frame wavelength coverage, and SED S/N. 

\begin{figure}[ht!]
    \includegraphics[width=485px]{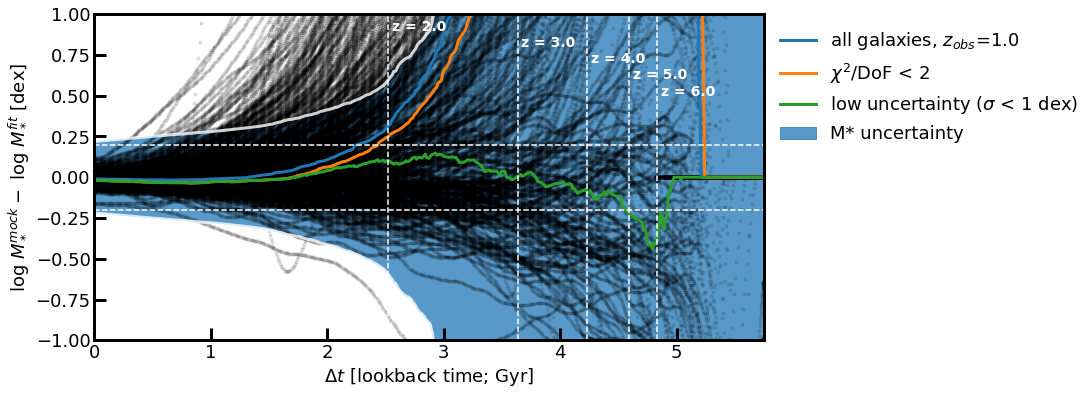}
    \includegraphics[width=485px]{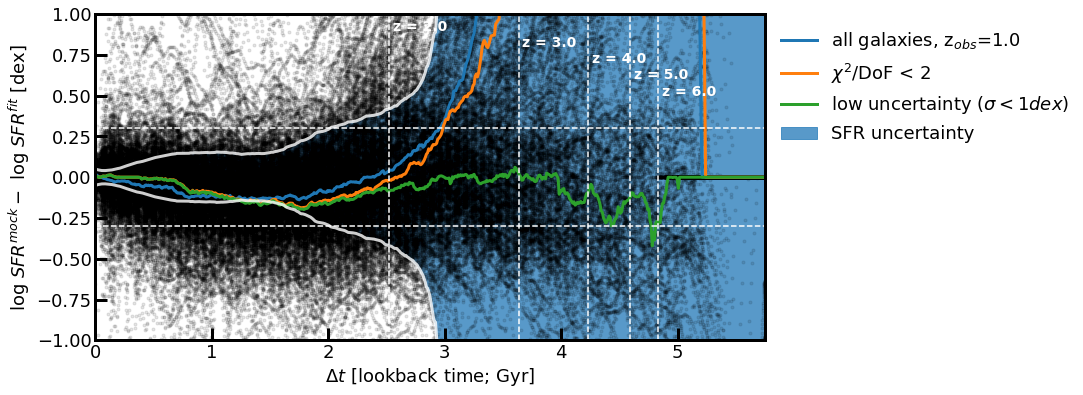}
    \caption{Reconstructed SFHs for an ensemble of mock galaxies \added{at $z\sim 1$} using the Dense Basis SED fitting method. Top panel shows the estimated bias (model - SED fitting estimate) and uncertainties (from SED fitting) in stellar mass as a function of lookback time, and the bottom panel shows the same for SFR. \newline
    In each plot, the \textbf{blue solid line} is pointwise median for all galaxies in the sample. This generally grows with time since our smooth SFHs can go to 0 while the mock SFHs generally go to some small nonzero value, which leads to a one-sided bias. However, since this bias would affect both SFR and M$_*$ identically, it simply shifts points along the diagonal, and shouldn't affect our estimates of slope and normalization. The \textbf{orange line} is pointwise median for the subsample of 'good fits' (method a) and the \textbf{green line} is pointwise median excluding contributions from galaxies which have large uncertainties ($\sigma \added{\log} (SFR(t),M_*(t)) > 1~dex$). (method c, using the uncertainties to avoid SFHs with biases). \newline
    The white horizontal line shows the average uncertainties on SFR ($\sim$0.3 dex) and M$_*$ ($\sim$0.2 dex), for comparison. The \textbf{white vertical lines} indicate redshifts instead of lookback times. (method b would be truncating our trajectories to where the orange line hits our tolerance bias of the white horizontal lines, here around $\Delta t \sim 2.5Gyr$.) \textbf{The blue shaded region + white solid line shows the uncertainty on SFR-M$_*$ estimate\added{s}, showing that the uncertainties closely follow the bias. Additionally, we see that for \replaced{all the samples}{the full ensemble of galaxies (sample (a))}, the uncertainties are larger than the bias and thus should not affect a statistical analysis that takes the uncertainties into account.}}
    \label{fig:validation_lbt_z1}
\end{figure}

Using a sample of mock galaxies at different redshifts and stellar masses, we span a range in S/N and wavelength coverage similar to what we see in the CANDELS data. For all of these galaxies we reconstruct the SFHs through SED fitting, along with uncertainties on \added{log} SFR(t) and \added{log} M$_*$(t) at each point in lookback time. We find that these estimated uncertainties at each point in lookback time trace of the bias in both \added{log} M$_*$ and \added{log} SFR, and allow us to restrict ourselves to a minimally biased subset of galaxies for fits at any redshift. This is seen in Figure \ref{fig:validation_lbt_z1}. 

We consider a few different options for reconstructing the high-z SFR-M$_*$ relation based on trajectory data from galaxies observed at later epochs (lower redshifts). 

\begin{itemize}
    \item \added{Method (a):} Use all available galaxies at lower redshifts / later epochs. 
    \item \added{Method (b):} Use a subset of available galaxies with bounds on how far back in lookback time  the galaxies are propagated, using SFHs from simulations to find the bounds.
    \item \added{Method (c):} 
    \deleted{Use a subset of available galaxies with thresholds on bias, which we find is traced well by the uncertainty on the fits.} 
    Using the simulations, we find that bias in estimating quantities like SFR, M$_*$ are traced closely by the \added{pointwise} uncertainties \added{on these quantities estimated during SED fitting. These uncertainties depend on factors such as the S/N of the SED, rest frame wavelength coverage, and degeneracies of the SFH with dust and metallicity. Since this criterion is based on the uncertainties in log SFR and log M$_*$ and not their actual values, this does not correspond to a selection in log sSFR}. \replaced{which we can use to limit}{We can thus reduce the bias by limiting} our analysis to a `high confidence’ subsample that avoids significant biases. \added{To ensure that the bias on M$_*$ and SFR are less than $0.2$ and $0.3$ dex respectively, we use a threshold of 1 dex on the combined uncertainties ($\sqrt{\sigma_{log M_* (t)}^2 + \sigma_{log SFR (t)}^2}$) to select galaxies for analysis at different epochs. We find that varying the threshold in the range of 0.5 to 2 dex does not significantly impact the analysis. This selection is possible for both the direct fits and trajectories since the SED fitting procedure estimates uncertainties on M$_*$ and SFR at each point in lookback time. For example, at $z=1$ we consider the uncertainties for all galaxies that are directly fit at $z\sim 1$ and exclude those with $\sqrt{\sigma_{log M_*}^2 + \sigma_{log SFR}^2} > 1 dex$. Similarly, for the trajectories we consider all the galaxies at $z<1$, and compute the uncertainties on their $M_*$ and $SFR$ after propagating them backwards in time to $z=1$ before applying our selection criterion. }

\end{itemize}

\begin{figure}[ht!]
    \includegraphics[width=500px]{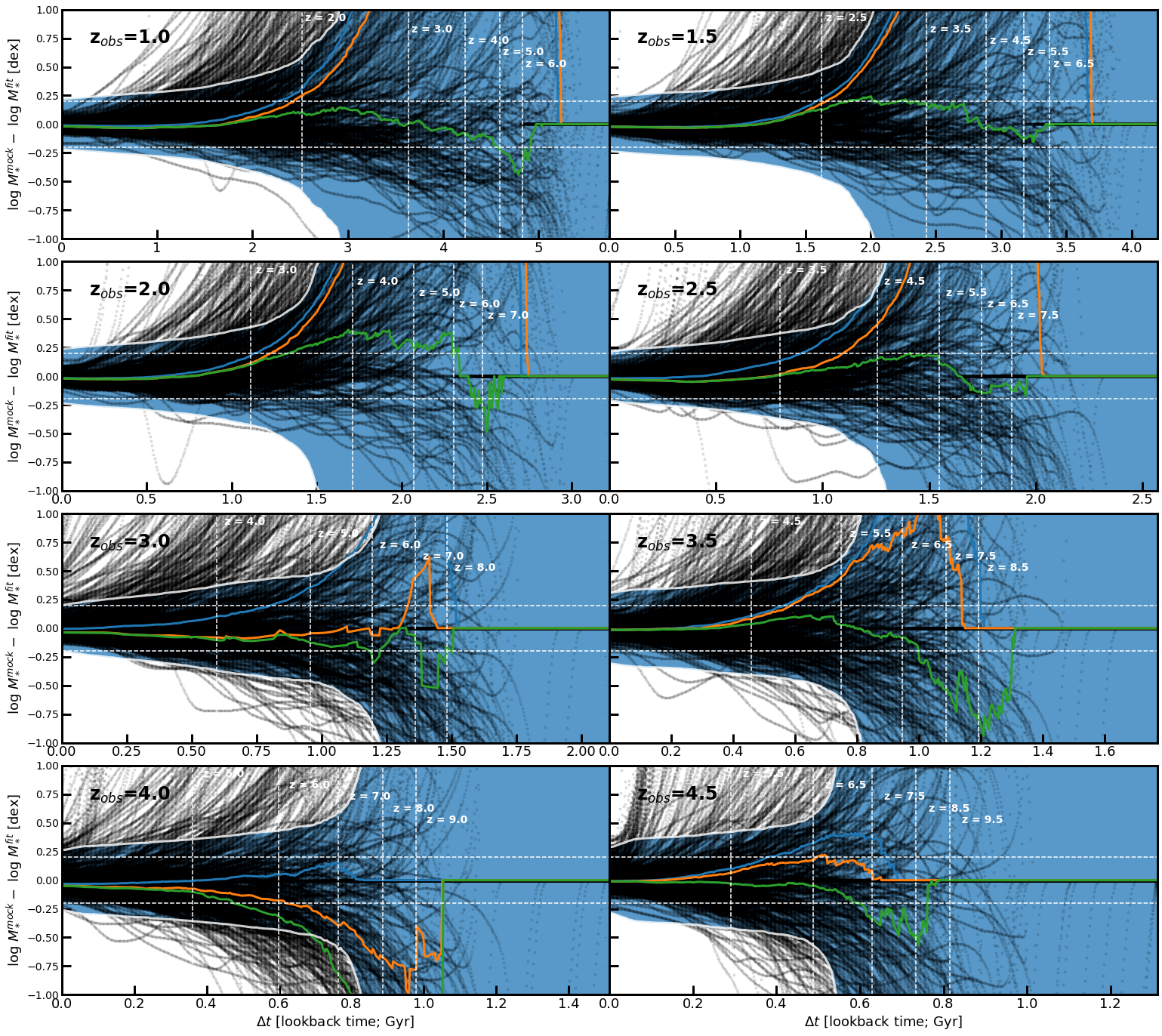}
    \caption{Same as figure \ref{fig:validation_lbt_z1} for Stellar Mass, repeated across a range of redshifts.}
    \label{fig:validation_lbt_mstar_allz}
\end{figure}

\begin{figure}[ht!]
    \includegraphics[width=500px]{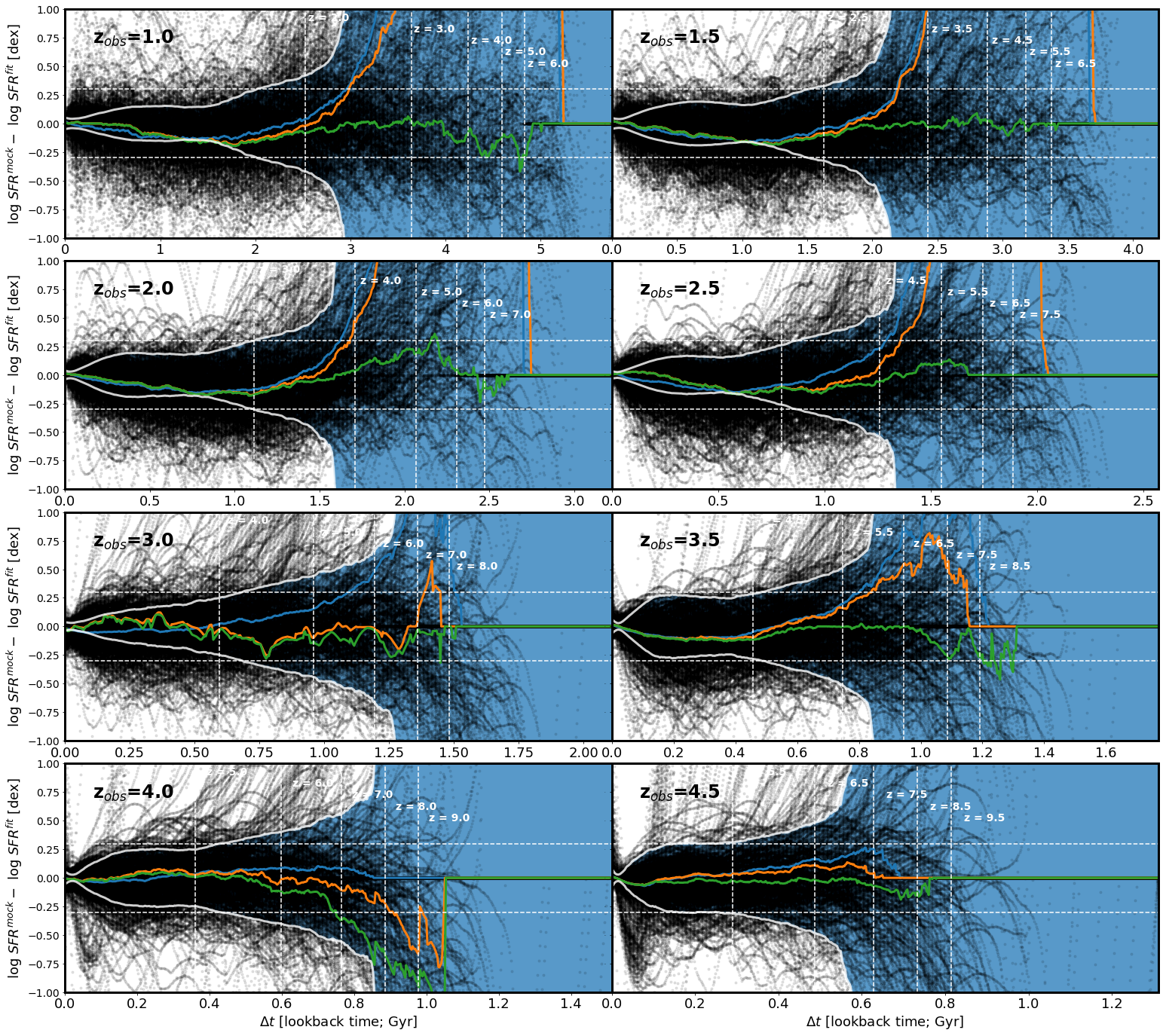}
    \caption{Same as figure \ref{fig:validation_lbt_z1} for Star Formation Rate, repeated across a range of redshifts.}
    \label{fig:validation_lbt_sfr_allz}
\end{figure}

These three approaches are highlighted in Figure \ref{fig:validation_lbt_z1}, where we show the error in estimating Stellar Mass and SFR for a sample from the simulations at $z\sim 1$ as a function of lookback time (i.e., trajectory run-time). In Figure \ref{fig:validation_lbt_mstar_allz} and Figure \ref{fig:validation_lbt_sfr_allz} we show this test run at a range of redshifts. We find that the dependence of the reconstruction on SED coverage or S/N is captured by the uncertainties, and thus using them as a tracer of the bias in our fits is an effective method that accomplishes both objectives: propagating robust SFR-M$_*$ trajectories farther back in time while avoiding samples with large uncertainties due to bad SED coverage, low S/N, or bad fits. \added{In Table \ref{table:delta_t_vs_z} we show the median amounts of time that galaxies in our actual analysis are propagated backwards along their trajectories to be included in the analysis at a given redshift of interest. While this increases as we go to higher redshifts, most galaxies are propagated backwards by only about $15-30\%$ of their full SFHs at any epoch.}

\begin{table}[ht!]
    \centering
    \begin{tabular}{c|c c c}
    \hline \hline
    $z_{analysis}$  &  \vtop{\hbox{\strut Median $\Delta t$}\hbox{\strut that galaxies are}\hbox{\strut propagated [Gyr]}} & \vtop{\hbox{\strut Median $z_{obs}$}\hbox{\strut [$t(z_{analysis})$- Median($\Delta t$)]}} & \vtop{\hbox{\strut $\Delta t$ between}\hbox{\strut $z=0.5$ and $z_{analysis}$}\hbox{\strut [Gyr]}} \\
    \hline
    1 & 0.8 & 0.82  & 2.7 \\
    2 & 0.8 & 1.58  & 5.1 \\
    3 & 1.3 & 1.89  & 6.3 \\
    4 & 1.4 & 2.21  & 6.9 \\
    5 & 2.1 & 1.98  & 7.3 \\
    6 & 2.2 & 2.07  & 7.5 \\
    \hline
    \end{tabular}
    \caption{Median amounts of time that sub-samples of low-uncertainty galaxies \added{(observed at $z_{obs}$)} are propagated along their redshifts to reach \added{desired} redshifts \replaced{of interest}{at which we have performed our analysis ($z_{analysis}$)}. We see that the amount of time grows as we go to higher redshifts, but is much less than the time interval between the lowest redshift and the redshifts of interest (the maximum amount of time a galaxy can be propagated in our current analysis). This indicates that most galaxies analyzed at a given redshift come from the vicinity of that redshift, rather than being propagated backwards all the way from $z\sim 1$.}
    \label{table:delta_t_vs_z}
\end{table}

\section{Robustness of trajectory - direct fit comparison to sample selection}
\label{app:kstest_diffsamples}

We use a KS test to compare the distributions of the distances from the best-fit SFR-M$_*$ correlation that we get from the direct fits to the distances we get from galaxies observed at later epochs and propagated backwards in time along their trajectories. The similarity of the two distributions in addition to our previous validation suggests that the reconstructions are robust and that the effects due to mergers do not significantly affect our analysis of quantities like the slope and normalization of the SFR-M$_*$ correlation. To ensure that the KS test is not affected by possible systematics arising from sample selection, we perform the test on a few different samples at each redshift: 

\begin{itemize}
    \item The full distribution of distances from both direct fits and trajectories, out \added{to} a distance of 0.4 dex from the best-fit correlation.
    \item The distribution of distances corresponding to the sample of galaxies with uncertainties $< 1 $ dex, out to a distance of 0.4 dex from the best-fit correlation.
    \item The distribution of distances corresponding to the sample of galaxies with uncertainties $< 1 $ dex, out to 1 $\times$ the observed scatter from the best-fit correlation.
    \item The distribution of distances corresponding to the sample of galaxies with uncertainties $< 1 $ dex restricted to the mass range where the direct fits have good statistics, out to 1 $\times$ the observed scatter from the best-fit correlation. To find the lower mass threshold, we find the minimum mass at for which the $median(M_*^{direct fits}) - median(m_*^{trajectories}) < 0.1$ dex .
\end{itemize}

The results of our KS test are consistent at all redshifts (p-value $> \alpha$), as summarized in table \ref{table:KStest_diffsets}. The significance level for each test \added{is} $\alpha = 0.05$. However, since we are performing a family of comparisons to test a single hypothesis, we need to control for the increased probability of false positives. To this end, we use adjusted significance levels of $\alpha' = \alpha/N =  0.05 /6 = 0.0083$ using a Bonferroni correction \citep{goeman2014multiple}, or $\alpha' = 1- (1-\alpha)^{1/N} = 1- (1-0.05)^{1/6} \approx 0.0085$ using the more conservative Sidak correction \citep{vsidak1967rectangular}. This choice of correction does not affect our results since our lowest p-value is $0.02$. Since our results remain consistent across this broad range of tests, we can not reject the null hypothesis that the two distributions are the same. While this does not completely rule out the possibility that the two distributions are different, the probability of this being the case is lower than $\alpha = 5\%$.

\begin{table}[]
    \centering
    \begin{tabular}{c|c c c c}
    \hline \hline
    \vtop{\hbox{\strut redshift of}\hbox{\strut analysis}} & \vtop{\hbox{\strut p-values}\hbox{\strut (full)}} & \vtop{\hbox{\strut p-values}\hbox{\strut (low uncert.)}} & \vtop{\hbox{\strut p-values}\hbox{\strut (low uncert.,}\hbox{\strut 1 $\times$ scatter)}} & \vtop{\hbox{\strut p-values}\hbox{\strut (low uncert.,}\hbox{\strut 1 $\times$ scatter,}\hbox{\strut M$_*$ threshold)}}  \\
    \hline
    1 & 0.14 & 0.85 & 0.91 & 0.18 \\
    2 & 0.04 & 0.10 & 0.11 & 0.15 \\
    3 & 0.07 & 0.02 & 0.15 & 0.80 \\
    4 & 0.88 & 0.67 & 0.56 & 0.30 \\
    5 & 0.56 & 0.20 & 0.50 & 0.03 \\
    6 & 0.52 & 0.98 & 0.82 & 0.56 \\
    \hline
    \end{tabular}
    \caption{P-values corresponding to the KS test comparing the distributions of distances from the best-fit SFR=M$_*$ correlation for the direct fits and trajectories. The first column (full) shows the p-values comparing the distribution across all $M_*$ within 0.4 dex of the best-fit line. The second column (low uncert.) repeats this analysis for the subsample of galaxies used for trajectories that have low uncertainties at the redshift of interest. The third column (low uncert., 1 $\times$ scatter) uses a threshold of 1 $\times$ the observed scatter at each redshift instead of a fixed threshold of 0.4 dex. The fourth column (low uncert., 1 $\times$ scatter, M$_*$ threshold) performs the KS test on a further reduced dataset restricted to the mass range where the direct fits have good statistics so that stellar mass effects don't enter into our comparison. The p-values for all these comparisons are $> \alpha $, indicating that the two distributions are not different to a statistically significant level.}
    \label{table:KStest_diffsets}
\end{table}

\section{Effects of UVJ pre-selection}
\label{app:UVJ_preselection}

The Star Formation Rates of actively star forming galaxies are tightly correlated with their Stellar Masses across a range of redshifts, with $\geq 68\%$ of such galaxies found within a narrow range of a single best-fit line. However, when galaxies enter periods of quiescence or undergo starbursts, they make excursions from this correlation. Improperly taking these galaxies into account (both by failing to exclude them or by being too rigorous in excluding them, thereby excluding some star forming galaxies as well) could lead to biases in our estimates of slope and normalization for the SFR-M$_*$ correlation. To mitigate this issue, we used an iterative robust fitting routine that excludes outliers \citep{holland1977robust}, which effectively re-calibrates to the data at each redshift slice to identify which points could be outliers. This avoids the use of pre-determined conditions for when galaxies are quiescent, since we find that our star-formation histories allow us to robustly distinguish between galaxies with low SFRs throughout their lifetime and galaxies that are experiencing a rapid fall in their SFR.

\begin{figure}[ht!]
    \centering
    \includegraphics[width=160px]{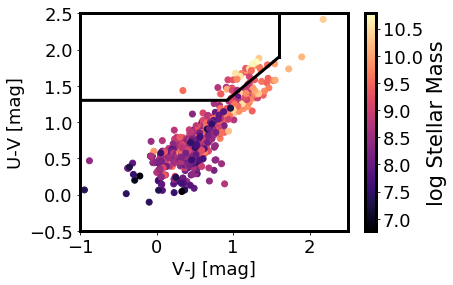}
    \includegraphics[width=160px]{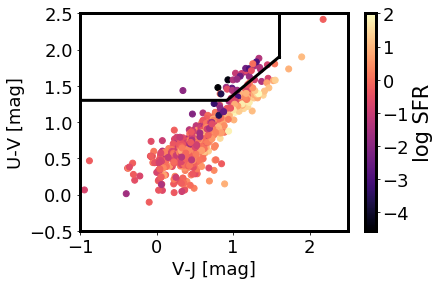}
    \includegraphics[width=160px]{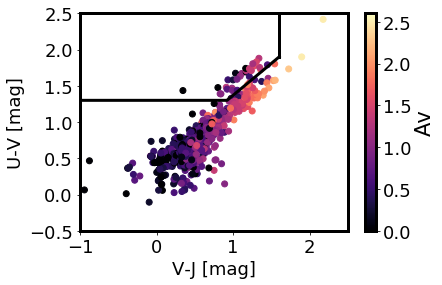}
    \caption{UVJ diagrams for the sample at $z\sim 1$. Rest frame U-V and V-J colors are computed through SED fitting, using the best-fit model template as a prior. The diagram shows the expected correlations with SFR and dust. }
    \label{fig:UVJ_diagrams}
\end{figure}

However, in order to better compare our results to literature that uses a pre-selection step to select star forming galaxies (see for example \citet{schreiber2015herschel}, where they use an optical selection criterion since even quiescent galaxies could still show residual IR emission due to a warm ISM), we adopt the UVJ selection criterion from \citet{williams2009detection}.
We use the rest-frame U-V and V-J colors derived by \citet{pacifici2016timing}. In brief, \citet{pacifici2016timing} use a large library of model SEDs to fit all available photometric data and derive median values and uncertainties of the rest-frame colors for each galaxy in the sample. The library is generated by combining the output of a semi-analytical model of galaxy formation with models of the stellar and gas emission and the attenuation by dust (see \citet{pacifici} for more details). 
The $z\sim 1$ sample with the selection criteria is shown in Figure \ref{fig:UVJ_diagrams}. 

Refitting the correlation to determine the slope and normalization does not significantly change our results, since most of the points excluded by the selection criterion would be classified as outliers by our algorithm. The fractional changes to slope and normalization at different redshifts are: -1.48 \%, -0.05 \%,  0.1 \% ,  0.78 \%,  0.4 \% , and  0.02 \%, while the fractional changes to normalization are: 0.01 ,  0.011,  0.025,  0.033,  0.011, and 0 dex at z=1,2,3,4,5,6 respectively.

\section{Using nonparameteric regression methods to quantify the SFR-M$_*$ correlation}
\label{app:nonparametric_fits_sfr_mstar}

In our analysis, we assume a linear relation between log SFR and log Stellar Mass and fit for its slope and normalization. However, it is not necessary that the correlation be linear. Indeed, \citet{schreiber2015herschel} and \citet{whitaker2014constraining} find that the relation flattens out at the high mass end. Using an F-test, we checked to see if a quadratic fit is statistically preferred over a linear one, and found this not to be the case. 

\begin{figure}[ht!]
    \centering
    \includegraphics[width=250px]{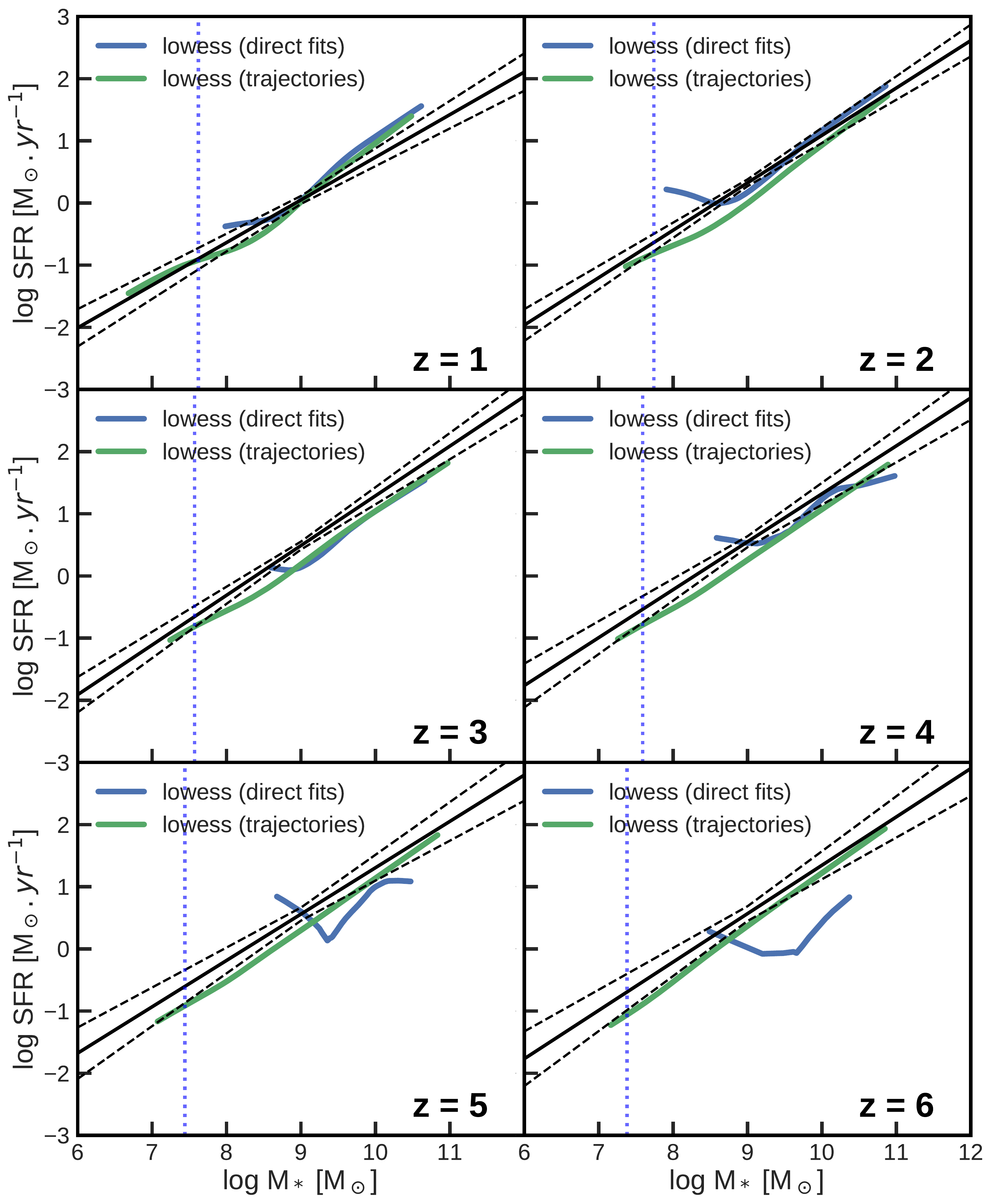}
    \includegraphics[width=250px]{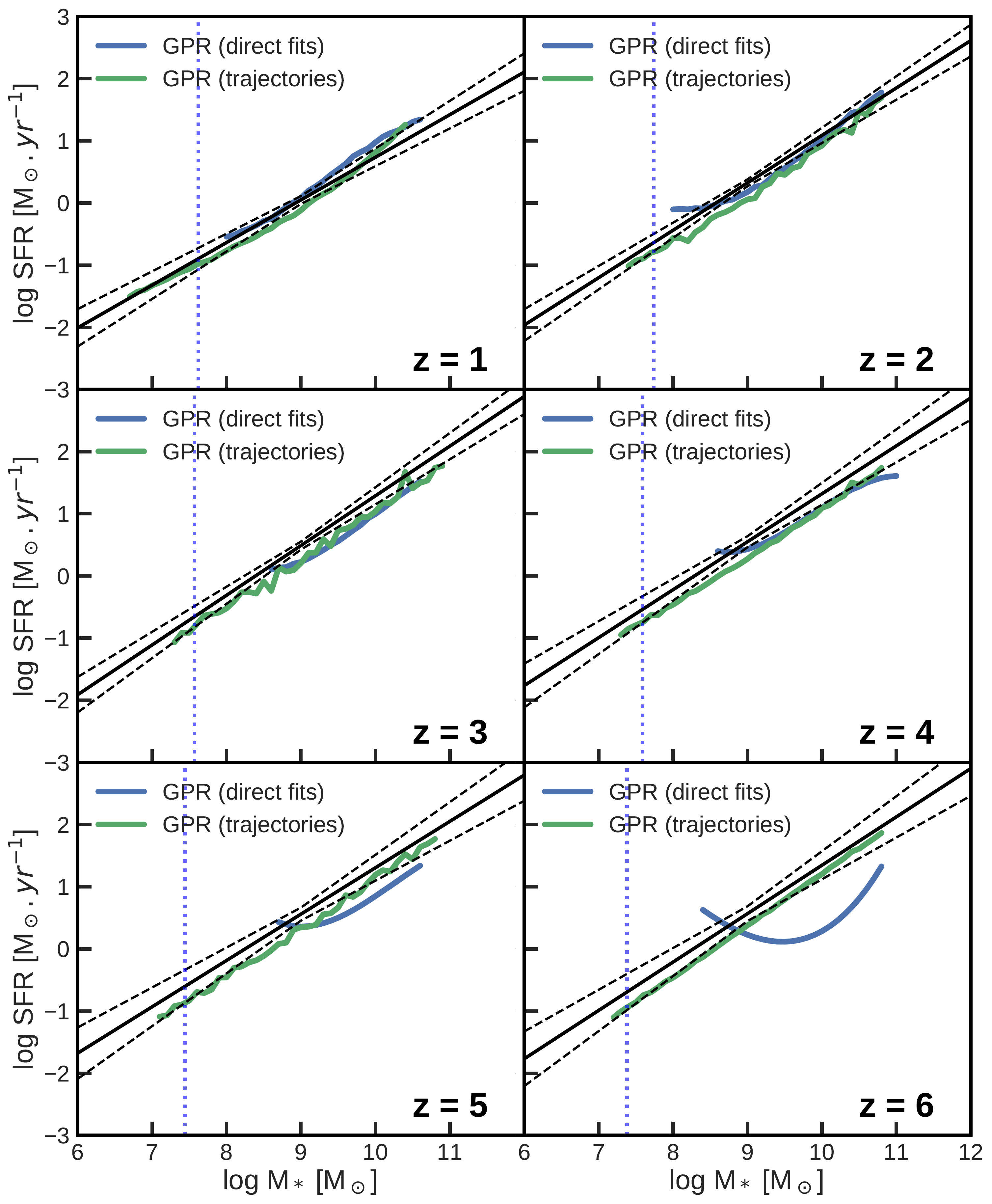}
    \caption{Nonparametric regression using Locally Weighted Scatterplot Smoothing (LOWESS) and Gaussian Process Regression (GPR) performed on the direct fits (blue lines) and trajectories (green lines) datasets at different redshifts. The curves are limited to the 5th to 95th percentile in Stellar Mass for each dataset with GPR, and from the 1st to 95th percentile for LOWESS. The curves corresponding to direct fits at $z>4$ are not reliable due to the small number of points available at those redshifts (see Figure \ref{fig:mainseq_allz}. \added{Dotted blue lines show the 10$^{th}$ percentile of the M$_*$ distributions at different epochs as reported in Table. \ref{table:lowest_masses_probed}.})}
    \label{fig:sfr_mstar_nonparametric}
\end{figure}

However, this is not enough to prove the linearity of the relationship. While some studies quantify the correlation between SFR and Stellar Mass by finding the effective SFR in bins of Stellar Mass, or even by binning perpendicular to the correlation, methods involving binning potentially suffer from effects due to bin size and the locations of bin centers. Here, we use LOWESS \citep{loess} a nonparametric regression technique that uses local weighting to create a smooth nonparametric estimate of the correlation. We also create a similar set of plots using Gaussian Process Regression (GPR) \citep{gp_book} to estimate the correlation at each point in $M_*$. From Figure \ref{fig:sfr_mstar_nonparametric}, we can see that the relation is indeed linear at the low mass end, closely matching our best-fit. While this alone isn't enough to state that the correlation is linear, it is certainly consistent within uncertainties with our best-fit linear relation.

From Figure \ref{fig:sfr_mstar_nonparametric} we also see that the direct fits and trajectories agree extremely well in the high-mass regime at redshifts \added{where} both sets have significant statistics. However, the high mass end has a slightly higher slope than the low-mass end at $z\sim 1-3$. From this, we conclude that when we fit the direct fits and trajectories, possible discrepancies may arise due to the different mass ranges over which they are being fit.

\listofchanges

\end{document}